\documentclass[twocolumn,preprintnumbers,amsmath,amssymb]{revtex4-1}
\usepackage{epsfig}
\usepackage{color}
\usepackage{amsmath}
\usepackage{float}
\usepackage{amsfonts}
\usepackage{amssymb}
\usepackage{multirow}
\usepackage{graphicx,stackengine}
\usepackage{subfig}
\usepackage[percent]{overpic}
\usepackage[english]{babel}
\usepackage{float}
\usepackage{array}
\usepackage{pst-plot}
\usepackage[version=4]{mhchem}
\usepackage{lipsum}
\usepackage{verbatim}
\usepackage{tikz}
\usetikzlibrary{calc,fadings}
\tikzfading[name=fade l,left color=transparent!100,right color=transparent!0]
\tikzfading[name=fade r,right color=transparent!100,left color=transparent!0]
\tikzfading[name=fade d,bottom color=transparent!100,top color=transparent!0]
\tikzfading[name=fade u,top color=transparent!100,bottom color=transparent!0]

\begin{document}
\title{Predictions of the interfacial free energy along the coexistence line from single-state calculations}

\author{Ignacio Sanchez-Burgos$^{1}$, Pablo Montero de Hijes$^{2,3}$, Eduardo Sanz$^{4}$, Carlos Vega$^{4}$ and Jorge R. Espinosa$^{1,4*}$}
\affiliation{
[1] Maxwell Centre, Cavendish Laboratory, Department of Physics, University of Cambridge, J J Thomson Avenue, Cambridge CB3 0HE, United Kingdom. \\
[2] Faculty of Physics, University of Vienna, 1090 Vienna, Austria \\
[3] Faculty of Earth Sciences, Geography and Astronomy, University of Vienna, 1090 Vienna, Austria \\ 
[4] Department of Physical-Chemistry, Universidad Complutense de Madrid, Av. Complutense s/n, 28040, Madrid, Spain. \\
* = To whom correspondence should be sent.
email: jorgerene@ucm.es}

\date{\today}

\begin{abstract}


\textcolor{black}{
The calculation of the interfacial free energy between two thermodynamic phases is crucial across various fields, including materials science, chemistry, and condensed matter physics. In this study, we apply an existing thermodynamic approach, the Gibbs-Cahn integration method, to determine the interfacial free energy under different coexistence conditions, relying on data from a single-state calculation at specified pressure and temperature. This approach developed by Laird \textit{et al.} [J. Chem. Phys. 131, 114110 (2009)] reduces computational demand and enhances efficiency compared to methods that require separate measurements at each thermodynamic state. The integration scheme computes the excess interfacial free energy using unbiased NVT simulations, where the two phases coexist, to provide input for the calculations. We apply this method to the Lennard-Jones and mW water models for liquid-solid interfaces, as well as the Lennard-Jones and TIP4P/2005 models for liquid-vapor interfaces. Our results demonstrate the accuracy and effectiveness of this integration route for estimating the interfacial free energy along a coexistence line.}

\end{abstract}
\maketitle


\section{Introduction}

The study of interfaces between different thermodynamic phases holds critical significance in various scientific and industrial domains. Interfaces are pivotal in understanding a wide array of physical phenomena, such as freezing \cite{zhang2016effect,espinosa2016interfacial}, nucleation \cite{sanchez2021fcc,hoose2012heterogeneous}, confinement \cite{rasaiah2008water}, as well as multiple technological \cite{asta2009solidification} and biological \cite{cochard2022rna,sanchez2021size} processes. Special focus has been devoted to the study of the solid-liquid interface \cite{bahadur2007surface,davidchack2003direct,davidchack2000direct,ambler2017solid,espinosa2014mold,espinosa2016ice,sanchez2021fcc,cheng2015solid,hoyt2001method,fernandez2012equilibrium}, however it is important to note that its comprehensive study is far from trivial due to its inherent complexity \cite{bahadur2007surface}.
\\

One of the pioneering contributions to the thermodynamic understanding of interfaces can be attributed to Gibbs. In his seminal work \cite{gibbs1928collected}, he introduced the concept of the interface between two different phases as a zero-width plane, often referred to as the "Gibbs' dividing surface". Gibbs associated this plane with the surplus of thermodynamic properties of the two phases once the interface is formed. Among the different parameters to characterize the properties of an interface, a key quantity is the interfacial free energy ($\gamma$), which quantifies the reversible work needed to generate an interface per unit of area. Later, Cahn reformulated this formalism using Cramer’s rule so that the excesses of extensive properties are expressed through differential coefficients given by
determinants that are composed of extensive properties of the interface \cite{johnson1979interfacial}. In such a way, there is no need for using the dividing surface.
\\

In the case of liquid-solid interfaces, determining the liquid-solid interfacial free energy ($\gamma_{ls}$) is not straightforward \textcolor{black}{generally due to the high computational cost associated to its calculation. As will be formulated later, $\gamma_{ls}$ cannot be directly measured from a liquid-solid coexisting system, and requires more sophisticated approaches to be calculated. One of the methods relies on} the measurement of crystal nucleation rates in supercooled fluids. Within the framework of the Classical Nucleation Theory (CNT) \cite{turnbull1950formation,volmer1926keimbildung,becker1935kinetische}, the estimation of $\gamma_{ls}$ is possible as a function of the saturation \cite{hale1996monte,piaggi2022homogeneous,sanchez2022homogeneous,espinosa2016seeding,espinosa2016time,lamas2021homogeneous,garaizar2022alternating,vazquez2009calculating}. 
\textcolor{black}{Another experimental indirect approach consists in measuring the angle a liquid droplet forms with a solid surface \cite{young1805iii,kwok1999contact}. However, issues such as evaporation, swelling, or surface imperfections affect accuracy, and the surface must be flat and chemically uniform. Deviations from these ideal conditions cause hysteresis between wetting and dewetting, which leads to non-unique contact angle measurements and making the Young equation unsuitable for calculating solid-liquid interfacial energy \cite{cha2019situ}. All the intricacies and necessary approximations} in experimental efforts to ascertain the liquid-solid interfacial free energy ($\gamma_{ls}$) \cite{na1994cluster,pruppacher1995new,ickes2015classical} underscore the growing significance of numerical approaches through molecular simulations to estimate this quantity. This growing importance is evident in the substantial body of research dedicated to developing and refining different computational methods. In fact, in the past two decades, a multitude of innovative methodologies have emerged in response to this need. Direct measurements of the interfacial free energy can be performed through various different techniques, such as Cleaving \cite{broughton1986molecular}, tethered Monte Carlo \cite{fernandez2012equilibrium}, Metadynamics \cite{angioletti2010solid}, Mold Integration \cite{espinosa2014mold,tejedor2024mold}, Capillary wave fluctuations \cite{hoyt2001method}, Gibbs-Cahn integration \cite{cahn1979interfacial} and other related thermodynamic integration schemes \cite{benjamin2014crystal,bultmann2020computation,leroy2009interfacial}. These techniques have demonstrated their capability to yield reliable estimates of the liquid-solid interfacial free energy for various crystallographic planes and a wide array of condensed matter systems \cite{ambler2017solid,davidchack2010hard,davidchack2003direct,sanchez2021fcc,asta2002calculation,sanchez2023direct,laird2009determination,laird2010calculation,espinosa2016interfacial,espinosa2015crystal,mu2006crystal,liu2013molecular,nair2012molecular,vazquez2009calculating}. It shall be noted that, regardless of their success in the calculation of such an elusive quantity, all existing techniques come with varying degrees of computational cost and implementation complexity. 
\\


\textcolor{black}{In this work, we apply the Gibbs-Cahn integration method \cite{laird2009determination,frolov2009solid,frolov2009temperature}, which allows for the determination of the interfacial free energy, $\gamma$, along the coexistence line, provided that the value of $\gamma$ is known for one thermodynamic state. By integrating the Gibbs-Cahn equation (a first-order differential equation), we can determine $\gamma$ at other points along the coexistence line. This bears some resemblance to the Gibbs-Helmholtz relation \cite{von1882thermodynamik}, which allows determining the change in chemical potential with temperature. With such method, the enthalpy is integrated, therefore it has to be know along the integration pathway. The accuracy of the Gibbs-Cahn method relies heavily on the precise determination of $\gamma$ at a single state point. Although we apply this approach for both liquid-solid and liquid-vapor interfaces, the latter case does not strictly require this methodology, as there are more direct ways of obtaining $\gamma$ for liquid-vapor systems. In what follows, we detail the methodology that enables the calculation of $\gamma$ along the coexistence line using straightforward unbiased simulations.}

\section{Methodology} \label{methods}

\textcolor{black}{In the Gibbs framework, the dividing surface separates the two bulk phases, and the surface excess quantities represent the deviation of the system’s real properties from the properties that would be obtained if the two bulk phases extended uniformly to the interface. In this framework, the excess energy,
entropy and number of particles per surface area for a system of two coexisting phases, 1 and 2, are given by} 

\begin{equation}
    e=(E-\rho_1^E V_1-\rho_2^EV_2)/A
    \label{excess_energy}
\end{equation}
\begin{equation}
    \eta=(S-\rho_1^S V_1-\rho_2^SV_2)/A
    \label{excess_entropy}
\end{equation}
\begin{equation}
    \Gamma=(N-\rho_1^N V_1-\rho_2^NV_2)/A
    \label{excess_number}
\end{equation}
where $e$, $\eta$, and $\Gamma$ are the surface excess energy, entropy, and number of particles per surface area, respectively, associated with the presence of an interface. $A$ is the interfacial area, $E$, $S$, and $N$ are the total energy, entropy, and number of particles of the system containing both phases in coexistence respectively. $\rho^E$, $\rho^S$, and $\rho^N$ are the bulk densities of energy, entropy, and number of particles respectively. $V$ is the volume of each phase denoted as 1 and 2. The values of $e$, $\eta$, and $\Gamma$ depend on the arbitrary choice of the dividing interface. Among all possibilities, the equimolar dividing surface is usually chosen and it is defined as that for which the value of $\Gamma$=0 (i.e. the surface density of excess molecules is zero).
\\

Alternatively, the formalism of Cahn \cite{cahn1978thermodynamics,sekerka2015thermal}, which builds on the work of Gibbs, uses a layer of finite thickness. In this case, Eqs. \ref{excess_energy}-\ref{excess_number} still apply although they correspond to the layer instead of the dividing surface. Cahn uses Cramer's rule to propose a compact thermodynamic description employing determinants of extensive properties \cite{sekerka2015thermal}. A particular choice
 of extensive variables makes this approach equivalent
 to the one of Gibbs for the equimolar dividing surface. 
By using Cahn's approach in such form, it was shown in Refs. \cite{frolov2009solid,frolov2009temperature,laird2009determination} that for liquid-solid interfaces, the interfacial free energy dependence along the coexistence line can be obtained from thermodynamic integration schemes. In particular, for a monocomponent system of cubic geometry under hydrostatic stress, the  expression 
 was suggested in Ref. \cite{laird2009determination}

\begin{equation} \black \begin{split}
    \left[\frac{d(\gamma_g/T)}{dT} \right]_{coex}= \\-\rho_s(T)^{-2/3}\left[\frac{e(T)}{T^2}+\frac{2\tau(T)}{3\rho_s(T)T}\left(\frac{d\rho_s}{dT}\right)_{coex}\right], \end{split}
    \label{integrationold}
\end{equation}

which can be integrated to yield

\begin{equation} \black
    \begin{split}
    \frac{\gamma_g}{T}-\frac{\gamma_{g,0}}{T_0}=
    \\
    -\int_{T_0}^T\rho_s(T)^{-2/3}\left[\frac{e(T)}{T^2}+\frac{2\tau(T)}{3\rho_s(T)T}\left(\frac{d\rho_s}{dT}\right)_{coex}\right]dT
    \end{split}
    \label{integration}
\end{equation}

where $\gamma_g$ is the interfacial free energy per surface atom: $\gamma_g=\rho_s^{-2/3}\gamma_{ls}$ (being $\rho_s$ the solid \textcolor{black}{number} density and $\gamma_{ls}$ the liquid-solid interfacial free energy) and $\tau$ is the interfacial stress, \textcolor{black}{which can be defined \cite{kirkwood1949statistical,alejandre1995molecular,de2006nature} as}

\begin{equation}
    \tau=\int_{-\infty}^{\infty}[p_N(x)-p_T(x)]dx=\frac{L_x}{2}(\overline{p}_N-\overline{p}_T)
    \label{taueq}
\end{equation}

where $L_x$ is the long side of the simulation box (perpendicular to the interfaces in our simulations), and $p_N$ and $p_T$ are the normal and tangential components of the pressure tensor with respect to the interfaces. \textcolor{black}{$p_T$ refers to the average over the two tangential directions. To improve statistics, the tangential component is averaged over the two tangential directions, which are equivalent to each other The factor 2 accounts for the existence of two interfaces when performing simulations under periodical boundary conditions.} Equation \ref{integration} is the key equation to solve here: it can be used to predict $\gamma$ at any temperature $T$ (and pressure along the coexistence line), once a value for the interfacial free energy ($\gamma_0$) at a given temperature ($T_0$) and pressure is calculated. We note that cubic geometry was assumed, which in the case of hexagonal ice is an approximation. Nevertheless, the interfacial free energy of cubic ice is very similar to that of hexagonal ice \cite{ambler2017solid}, so we do not expect this approximation to have any significant effect. \textcolor{black}{Notice that all properties of  the integrand of Eq. \ref{integration} are computed along the liquid-solid coexistence curve (so that when T changes p changes to the corresponding value of the coexistence curve). }A requirement to solve Eq. \ref{integration} is to know how the interfacial excess energy ($e$) depends on $T$, which can be easily computed through Eq. \ref{excess_energy}. The energy densities for the bulk phases can be computed by performing NpT simulations of the bulk phases 1 and 2 separately. The total energy ($E$) is obtained from a Direct Coexistence (DC) simulation \cite{Ladd1977Triple-pointSystem}. DC simulations are performed in the canonical ensemble, and in such simulations, the two coexisting phases are in direct contact forming two interfaces separating them. The volume occupied by each coexisting phase ($V_1$ and $V_2$) is easily calculated knowing the coexisting density of each phase (previously calculated through NpT simulations). To do so, we take into account that the layer in the formalism of Cahn is defined by flat planes and that we choose to use it in the form that is equivalent to setting the Gibbs dividing surface at $\Gamma = 0$ (see Eq.\ref{excess_number}) to obtain $V_1$ and $V_2$  by solving

\begin{equation}
    \left. V=V_1+V_2 \atop N=\rho_1^NV_1+\rho_2^NV_2 \right\}
    \label{sistema}
\end{equation}

where $N$ and $V$ are input simulation parameters. \textcolor{black}{We stress the fact that it is essential to ensure in DC simulations the solid crystal lattice is not strained under the imposed temperature and simulation box dimensions (i.e., remains at the equilibrium solid density).}
\\

Lastly, we need to obtain $\tau$ (Eq. \ref{taueq}) in order to solve Eq. \ref{integration}. This is another straightforward measurement, directly inferred from the DC simulations, since the pressure tensor can be directly calculated during the simulation. In summary, the present methodology requires performing a small amount of unbiased simulations (NVT DC simulations + bulk crystal and bulk fluid simulations) to obtain the dependence of the interfacial excess energy ($e$) and the excess interfacial stress ($\tau$)
with temperature and, once known $\gamma_{ls}$ for a given coexistence point, we can predict the liquid-solid interfacial free energy along the coexistence line. \\

As discussed in Ref. \cite{di2020shuttleworth}, the described framework is useful for interfaces containing solids. For fluid-fluid interfaces, the 
same formalism applies but becomes simpler. 
In particular, 
 Eq. \ref{integrationold} can be rewritten (see SM) as

 \begin{equation}
    \left[\frac{d(\gamma/T)}{dT} \right]_{coex}=-\frac{e}{T^2} - (\tau - \gamma)\frac{2}{3\rho_sT}\left(\frac{d\rho_s}{dT}\right)_{coex}.
    \label{integration2}
\end{equation}
 
 However, for fluid-fluid interfaces $\tau = \gamma$, which makes the second term in the right-hand-side zero.  While solids may present strained states  
like the stretching or compression of the lattice, fluids do not do so due to their capability to reorganize their structure upon temperature/pressure variations in a given simulation box. That is, for a given thermodynamic state the fluid-fluid interfaces adapt to reach always the same density but solids can be forced to a different density if their lattice is strained. Another way to see this is that the density
   of a fluid phase is not a variable of the state of the interfacial system, whereas a solid, which is constrained by its structure, requires this variable to describe the thermodynamic state. Therefore, for liquid-vapor interfaces, the interfacial free energy along the coexistence line can be obtained as

\begin{equation} \black
    \int _{1/T_0}^{1/T} e\left(\frac{1}{T}\right)\ d\left(\frac{1}{T}\right)=\int _{\gamma_{lv,0}/T_0}^{\gamma_{lv}/T}d\left(\frac{\gamma_{lv}}{T}\right)
    \label{lv1}
\end{equation}

which after integrating the expression results in

\begin{equation} \black
    \frac{\gamma_{lv}}{T}=\frac{\gamma_{lv,0}}{T_0}+\int _{1/T_0}^{1/T} e\left(\frac{1}{T}\right)
    d\left(\frac{1}{T}\right)
    \label{integrallv}
\end{equation}

where $\gamma_{lv}$ refers to the liquid-vapor interfacial free energy. e(1/T) denotes de dependence on 1/T of the excess energy.
\\

In this work, we exploit the thermodynamic integration approach for computing $\gamma$ along the liquid-solid coexistence line of the Lennard-Jones potential and the mW water model, and for the liquid-vapor coexistence line of the Lennard-Jones and TIP4P/2005 water model.

\section{Simulation details}

We employ the Broughton and Gilmer (BG) truncated Lennard-Jones potential  \cite{broughton1986molecular}, the atomistic non-polarizable TIP4P/2005 water model \cite{abascal2005general}, and the mW \cite{molinero2009water} coarse-grained water model. The potentials for all these models can be found in the Supplementary Material (Section SI). Simulations were performed with the Large-scale
Atomic/Molecular Massively Parallel Simulator (LAMMPS) \cite{Plimpton1995FastDynamics} Molecular Dynamics package for the case of the Lennard Jones and mW potentials, and the GROMACS
Molecular Dynamics package \cite{bekker1993gromacs} (version 4.6.7) for the TIP4P/2005 model.
\\

For the Lennard-Jones BG potential, we fix $\epsilon=1, \sigma=1, m=1$, and use reduced units, indicated with the symbol $^*$ (further details can be found in the SM), where $T^*=k_BT/\epsilon$. We perform DC simulations in the NVT ensemble, keeping the temperature constant with the Nosé-Hoover thermostat \cite{nosethermo,hooverthermo,hoover1986constant}. For the simulations in the NpT ensemble (i.e., bulk liquid or bulk crystal systems), we fix the pressure with the Nosé-Hoover barostat \cite{nosethermo,hoover1986constant,hooverthermo}. The relaxation times are 1 and 10 (in reduced time units) for the thermostat and barostat, respectively. We integrate the equations of motion using the velocity-Verlet integrator. The simulation time step chosen is 0.001 in reduced time units.
We carry out Mold Integration calculations \cite{espinosa2014mold} using the LAMMPS Mold package \cite{tejedor2024mold} developed by us.
\\

For the simulations of the mW model in LAMMPS, we set the relaxation time of the thermostat and barostat at 0.1 and 0.5 ps respectively. The simulation time step chosen is 5 fs. The forces calculated using the mW model vanish at a
cut-off distance of $r_c=a\sigma=1.8\ \sigma$, being $\sigma=$2.3925 \r{A} the water molecular diameter as proposed in Ref. \cite{molinero2009water}. The same barostat and thermostat as in the Lennard-Jones BG simulations were used. 
\\

For the TIP4P/2005 simulations, we employ GROMACS. The v-rescale (Bussi-Donadio-Parrinello) thermostat \cite{bussi2007canonical} is used in both NVT and NpT simulations. For NpT simulations, the Parrinello-Rahman barostat \cite{parrinello1981polymorphic} is also used to keep pressure constant. We integrate the equations of motion using the Leap-Frog integrator \cite{hockney1974quiet}. The simulation timestep chosen is 0.002 ps, and the thermostat and barostat relaxation times are 2 and 0.75 ps, respectively. For the TIP4P/2005 potential, we set the cut-off of both dispersive interactions and the real part of the electrostatic interactions at 1.3 nm, using a switch radius at 1.2 nm \cite{p2023water}, and no long-range dispersion corrections for energy and pressure are applied. Long-range Coulombic interactions were treated with the Particle-Mesh Ewald (PME) solver \cite{darden1993effect,essmann1995smooth}. We keep the O-H bond length (0.9572 \r{A}) and H-O-H angle (104.52 º) values constant with the LINCS algorithm implemented in GROMACS \cite{hess1997lincs}. 

\section{Results and discussion}

\subsection{Thermodynamic integration of $\gamma$ along the liquid-solid coexistence line of Lennard-Jones particles}

\begin{figure}
    \centering
    (a) Lennard-Jones BG \\
    \includegraphics[width=\linewidth]{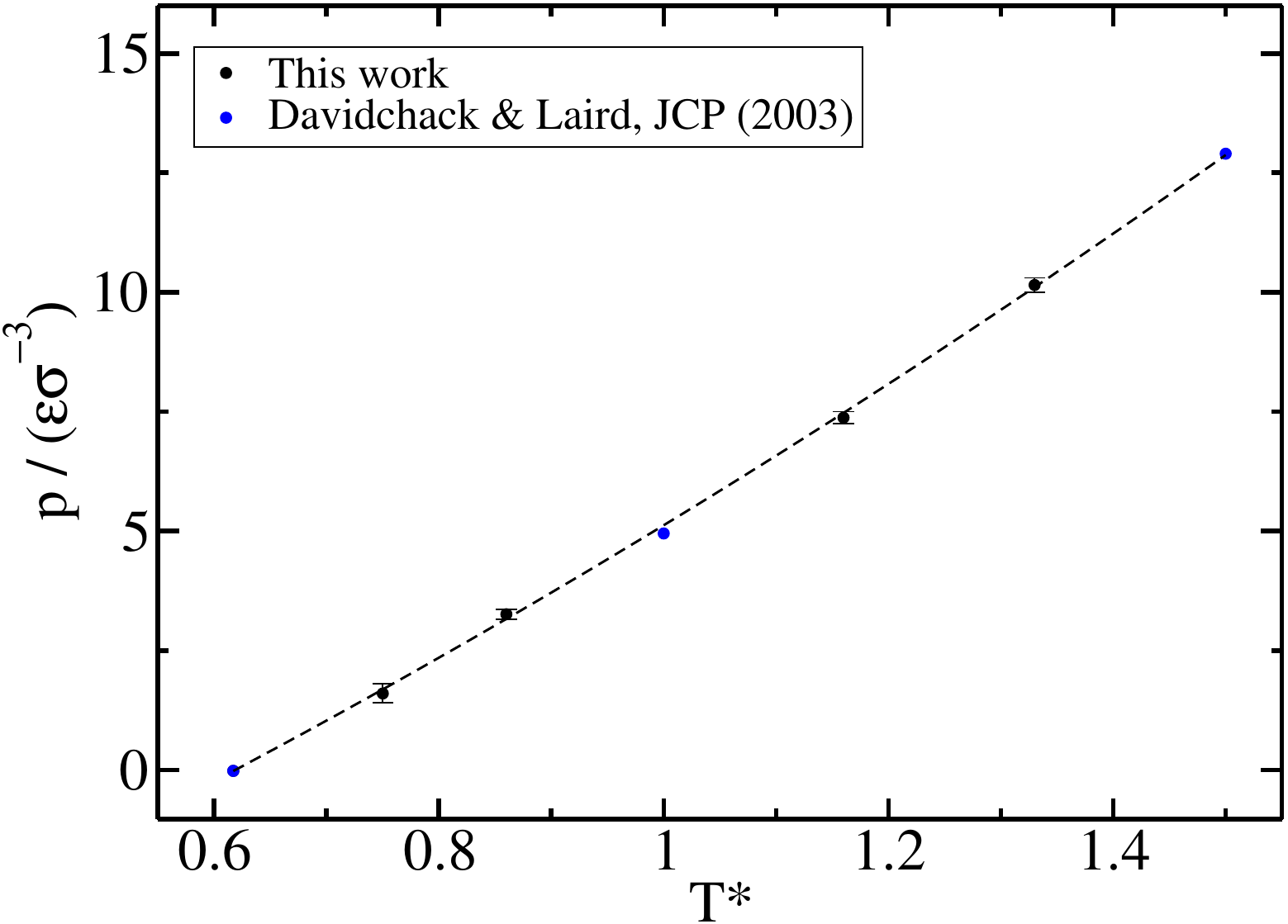} \\
    (b) mW \\
    \includegraphics[width=\linewidth]{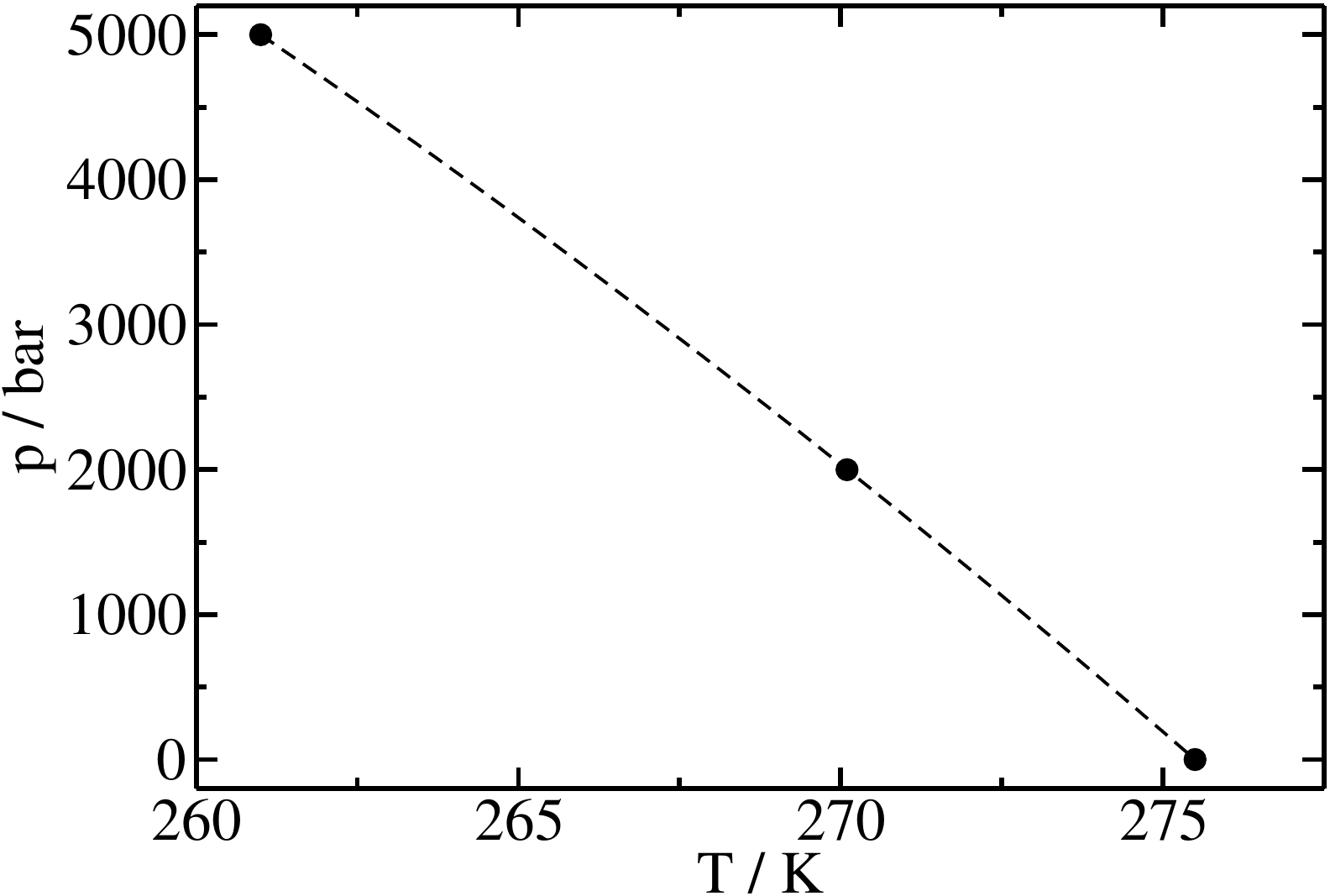}
    \caption{Liquid-solid coexistence line in the pressure-temperature plane for: (a) the Lennard-Jones system; and (b) the mW water model.}
    \label{coexostenceslines}
\end{figure}

\begin{figure*}[ht!]
    \centering
    \begin{tabular}{cc}
        (a) & (b) \\
        \includegraphics[width=0.5\linewidth]{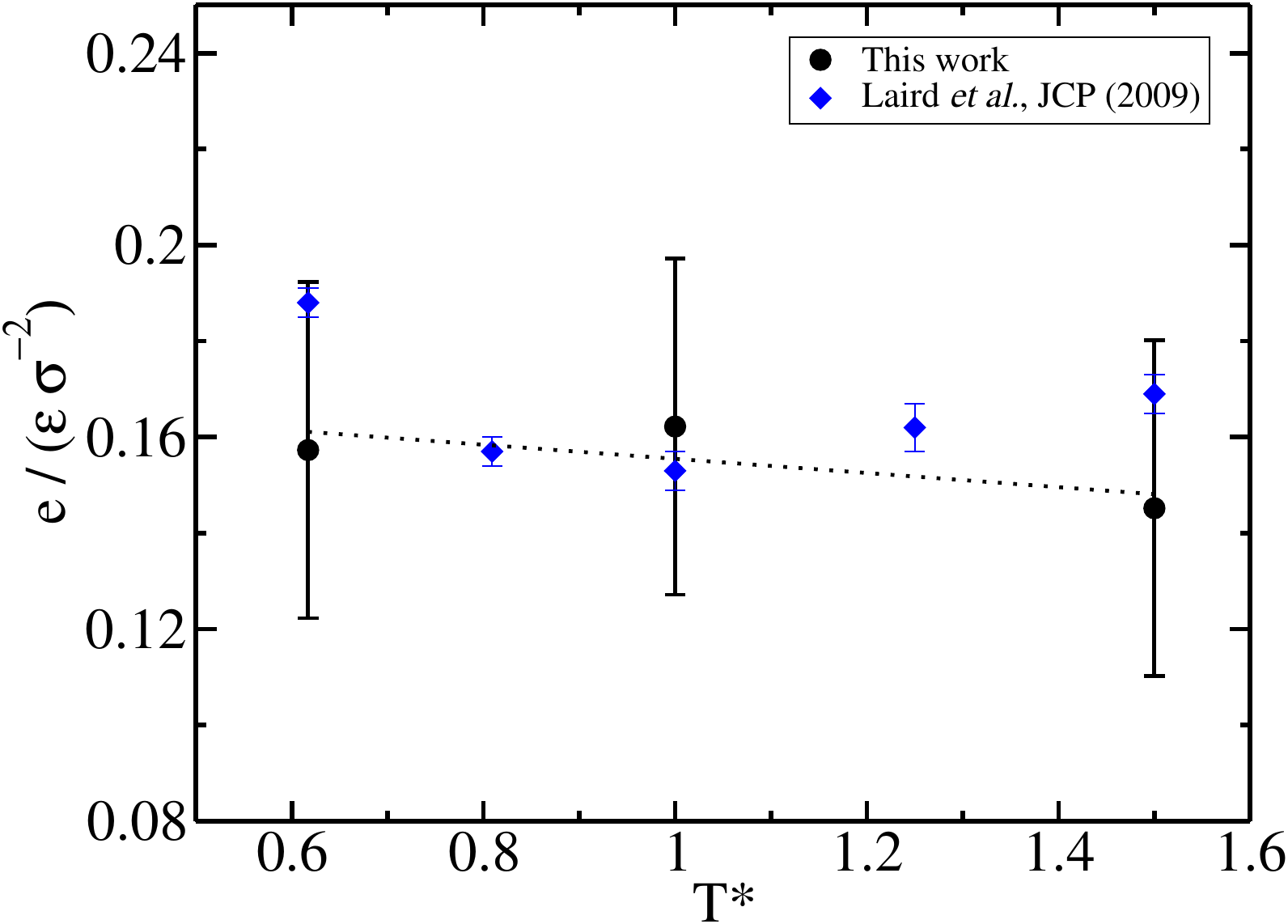} & \includegraphics[width=0.5\linewidth]{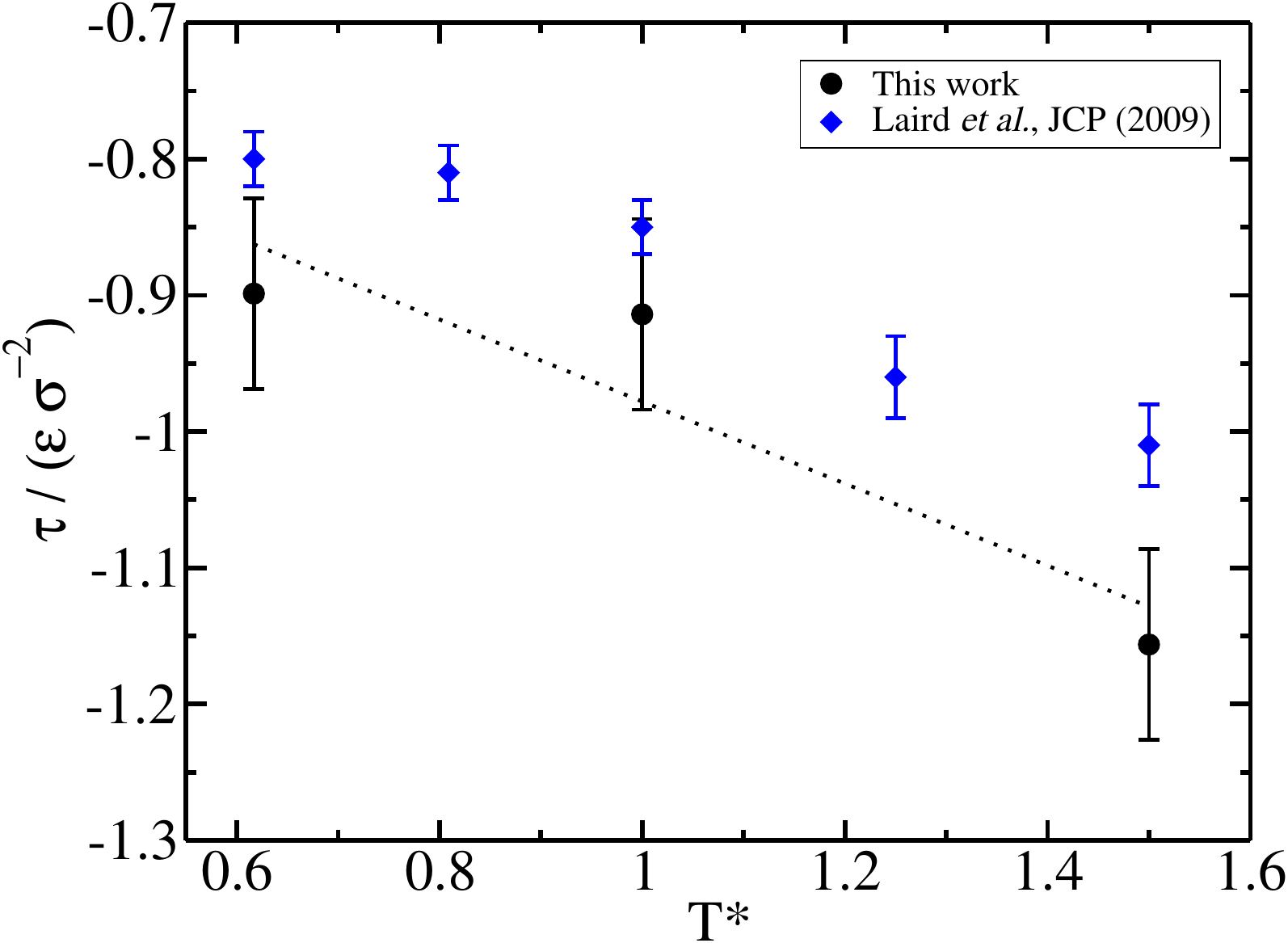} \\
        (c) & (d) \\
        \includegraphics[width=0.5\linewidth]{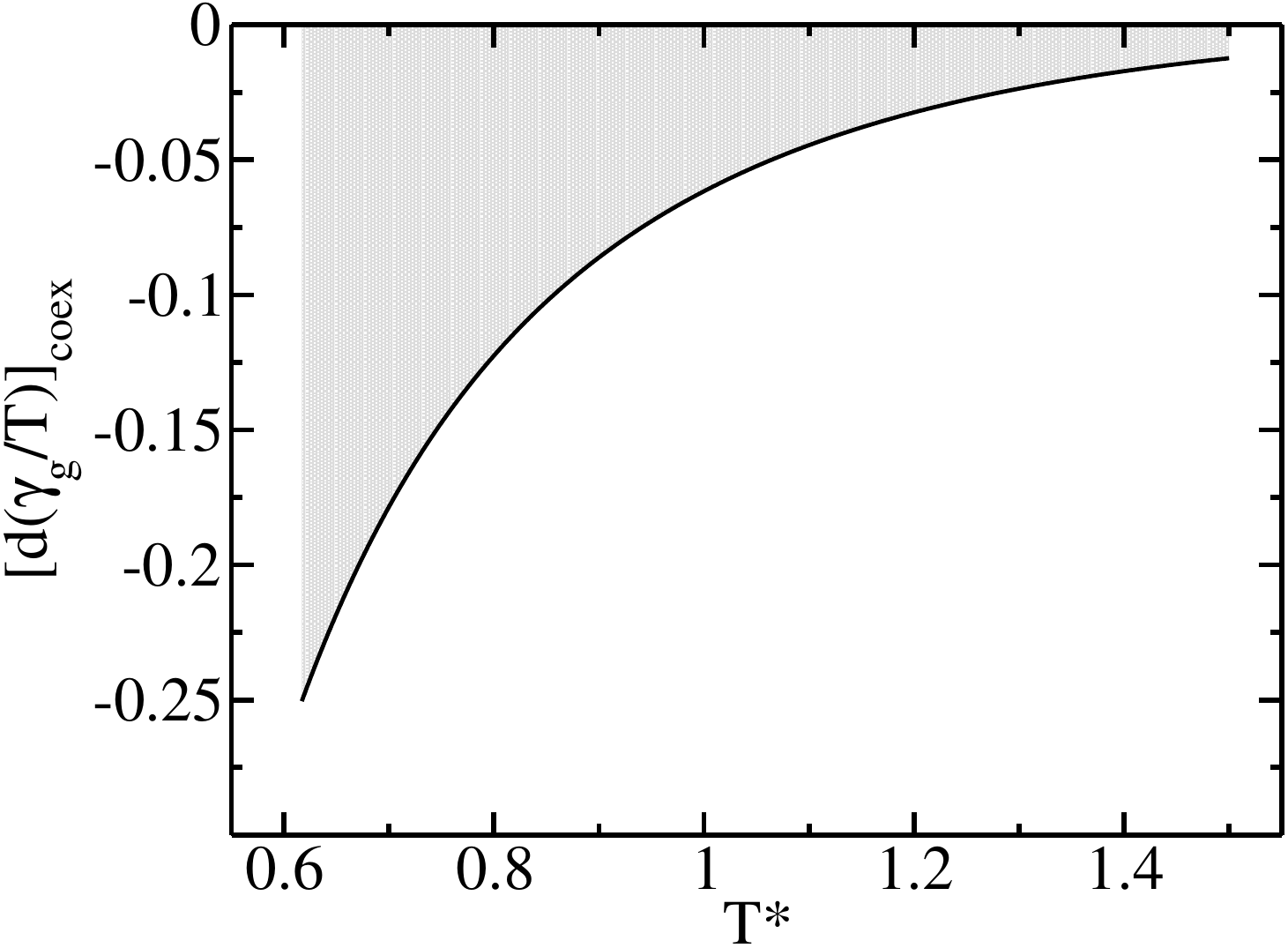} & \includegraphics[width=0.5\linewidth]{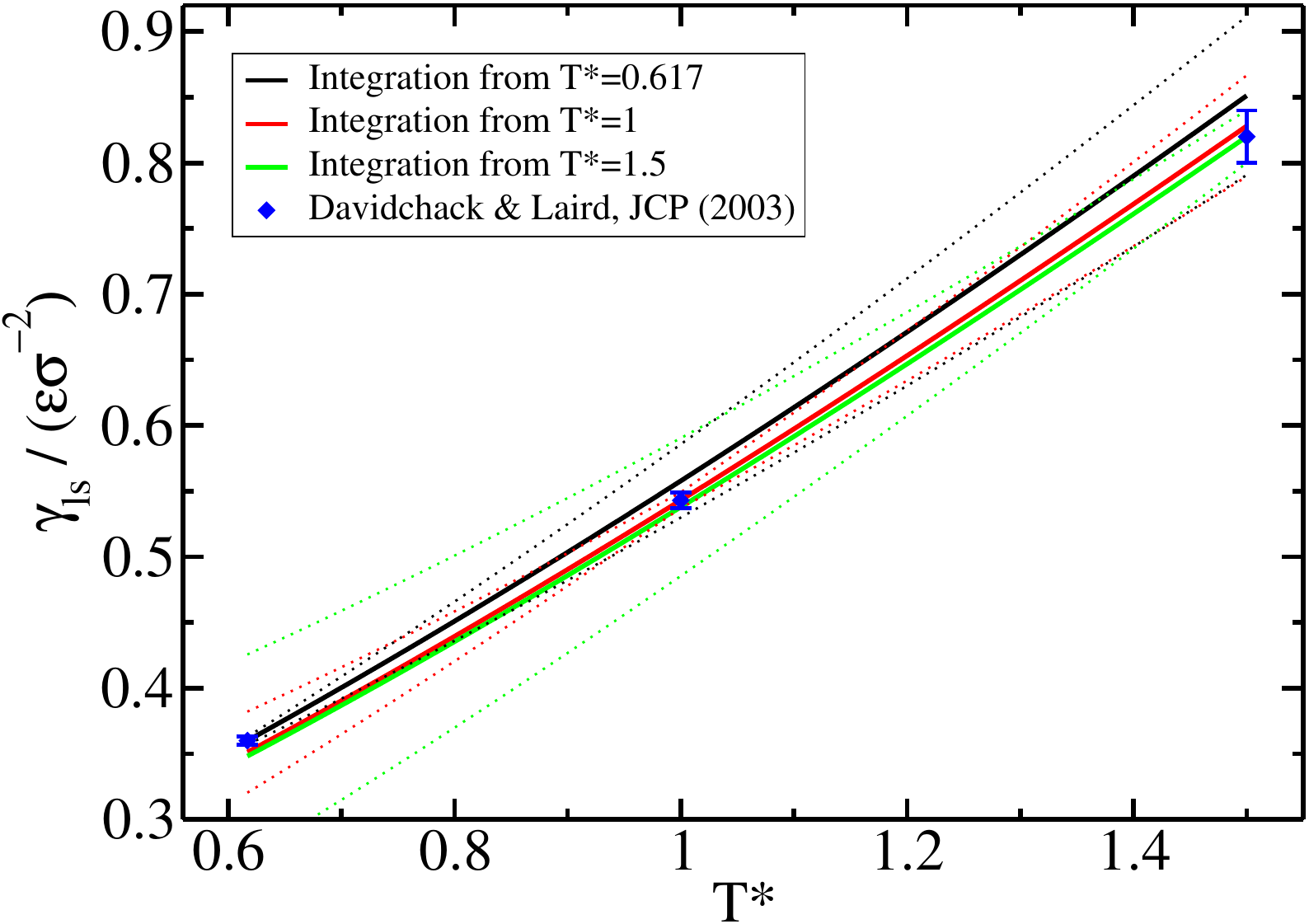}
    \end{tabular}
    \caption{Gibbs-Cahn integration of $\gamma_{ls}$ for the BG Lennard-Jones fcc (110) plane: (a) Excess interfacial energy as a function of temperature along with results from Laird \textit{et al.} \cite{laird2009determination}. The black dashed line represents a linear fit to our data: $e=0.1702-0.01471T^*$ (where $T^*=k_BT/\epsilon$). (b) Excess interfacial stress as a function of temperature, along with results from Laird \textit{et al.} \cite{laird2009determination}. The black dashed line represents a linear fit to our data: $\tau=-0.6767-0.3012T^*$. (c) Slope of $\gamma_g/T$ as a function of temperature for the (110) plane along the melting line. The shaded region represents the integral of such function along temperature. (d)
    Predicted interfacial free energy  as a function of $T^*$ along the coexistence line for (110) plane. The continuous lines indicate the estimations obtained from Eq. \ref{integration}, using the $\gamma_{ls}$ values from Ref. \cite{davidchack2003direct} (blue diamonds). Dotted lines indicate the uncertainty limits for each of the estimations from the different states as indicated in the legend.}
    \label{fig1}
\end{figure*}

We begin by applying the Gibbs-Cahn thermodynamic integration methodology to the study of $\gamma_{ls}$ along the liquid-solid (face centered cubic; fcc) coexistence line of the Lennard-Jones BG potential. The coexistence line in the $p^*-T^*$ plane indicates the coexistence conditions between the liquid and solid phases. Such curve is shown in Figure \ref{coexostenceslines} for the Lennard-Jones BG and mW models, which will be studied later. In order to determine the coexistence conditions, we performed Direct Coexistence simulations in the NpT ensemble. Further details on how to obtain the coexistence conditions is detailed in the Supplementary Material Section SII.
\\

We first illustrate the thermodynamic integration methodology for the (110) plane of the fcc phase in the BG model. We prepare a NVT DC simulation containing $\sim$10000 particles, and we obtain $E$ (to later use it in Eq. \ref{excess_energy}), as well as $\tau$ from Eq. \ref{taueq}. It is crucial to set the correct box dimensions for NVT DC simulations. Since the coexistence density for the fcc solid is known \cite{broughton1986molecular}, we prepare simulation boxes that perfectly accommodate the solid at its equilibrium lattice paramters along the directions tangential to the interface at the chosen temperature---otherwise the solid phase will be stressed and hence its chemical potential will be higher than that at the coexisting conditions. We prepare simulation boxes that initially contain the same amount of solid and liquid particles, and fix total density around the average between the equilibrium densities of the respective bulk solid and liquid phases at coexistence conditions. In principle, the choice of the amount of inserted particles of each phase is not relevant since, as long as the global density is between $\rho_s$ and $\rho_l$, the system will spontaneously equilibrate (and the pressure along the direction perpendicular to the interface matches the coexistence pressure) in order to accommodate the two coexisting phases in equilibrium, regardless of our choice for the initial configuration. Then, in parallel, the bulk simulations are performed for both the liquid and solid phases at the corresponding coexistence conditions using NpT simulations (with systems sizes of $\sim$10000 particles), from where we obtain $\rho_1^E$ and $\rho_2^E$. With such information, we obtain the excess interfacial free energy, $e$, by solving Eq. \ref{excess_energy}, where A is directly measured from the DC simulation box length ($A=2*L_y*L_z$) and $V_1$ and $V_2$ are obtained from Eq. \ref{sistema}. 
\textcolor{black}{For the sake of computational efficiency we use system sizes about ten times smaller than those employed in Ref. \cite{laird2009determination}. In spite of the large  system size difference, our results are consistent with those of Ref. \cite{laird2009determination}, suggesting the absence of finite-size effects in our simulations (we note that the shortest dimension in our DC simulations fits at least 12 unit cells, spanning more than five time the cutoff distance). Moreover, using a different technique (Metadynamics), in Ref. \cite{angioletti2010solid} finite size effects were analysed in calculations of liquid-solid interfacial free energies and reliable results were obtained for systems even smaller than ours.}
\\

In Figure \ref{fig1}(a) we plot $e$ against $T$ and its linear fit, along with the previously calculated values from Ref. \cite{laird2009determination}. Note that for each temperature the system is at a different pressure (given by the corresponding coexistence pressure at the imposed temperature). Moreover in Figure \ref{fig1}(b) we also plot $\tau$ against $T$, once again comparing it to the results from Laird \textit{et al.} \cite{laird2009determination} for the same crystal orientation and Lennard-Jones BG potential. \textcolor{black}{Unless otherwise stated, we perform linear regression due to 1) the fact that overall, the data we report in this Figure and the following ones may follow this trend; 2) the difficulty to assign a higher degree polynomial fit to data in which we use a limited amount of points (3 in this case); Our choice of using linear fits is also justified by the fact that we find a good agreement between the $\gamma_{ls}$'s we obtain through 
integration and those computed with direct estimates.} \\

We now proceed to evaluate Eq. \ref{integration}. In Figure \ref{fig1}(c) we show the integrand of the right hand side term of Eq. \ref{integration} as a function of temperature (and implicitly of pressure through the variation along the coexistence line). 
Here, we apply the thermodynamic integration methodology discussed in Section \ref{methods}. As an extension of previous work \cite{baidakov2014surface,laird2009determination}, we have obtained fitted expressions for $e$, $\tau$ and $\rho_s$ with $T$. Therefore, we continuously evaluate the integrand of Eq. \ref{integration} and solve the equation, which yields a continuous prediction of the integration, while in the original proposed method, Eq. \ref{integration} was discretely evaluated for the states at which DC and bulk simulations were performed, and the integrand was calculated by means of the trapezoidal integration method. Here, we fit the behaviour of $e$, $\tau$ and $\rho_s$ with $T$ through fits \textcolor{black}{(linear for $e$ and $\tau$ and 2nd degree for $\rho_s$ unless stated otherwise), as opposed to only evaluating the integrand at the states at which we have directly computed $e$, $\tau$ and $\rho_s$. This allows us to easily obtain continuous predictions of $\gamma_{ls}$ with $T$ in a simple manner, as opposed to integrating using the trapezoidal rule integration method}. The dependence of $\rho_s$ against $T^*$ is required \textcolor{black}{and we obtain an analytical expression for it (second order polynomial fit, Figure S3)}. The shaded area depicted in Figure \ref{fig1}(c) represents the integrated Eq. \ref{integration}, and using a \textcolor{black}{integration starting} value of the interfacial free energy for a given state point, we obtain the behaviour of $\gamma_{ls}$ against temperature under coexistence conditions (Figure \ref{fig1}(d)). \\

We use the reported $\gamma_{ls}$ \textcolor{black}{values from Ref. \cite{davidchack2003direct} as starting points ($\gamma_{ls,0}$ and $T_0$) for thermodynamic }integration. In Figure \ref{fig1}(d) we show the predicted curves for $\gamma_{ls}$. The estimations from the Gibbs-Cahn integration (independently of the starting state point) are in excellent agreement with the individually computed values of $\gamma_{ls}$ at different temperatures\textcolor{black}{, since within the error bars of both the direct estimates and the Gibbs-Cahn predictions, the values obtained from both approaches match. We estimate the uncertainty bounds by combining two sources of error: the error in determining $\gamma_{ls}$ at the starting integration point (which comes from the technique used to directly determine $\gamma_{ls}$), and the error in $e$ and $\tau$, which propagates with temperature. For this second component, we calculate the maximum error in $e$ and $\tau$ at the furthest integration point and linearly interpolate this error for intermediate points along the thermodynamic integration pathway. We account for the fact that the error in $e$ and $\tau$ is zero at the lower bound of the integration, i.e., at $T = T_0$. The error in $e$ and $\tau$ is obtained from block analysis, in which we divide the simulation data into uncorrelated blocks that are averaged each of them. Then, the error is obtained as the standard deviation of the different blocks over the square root of the number of blocks. This is later multiplied by a t-student value for a 95\% confidence level to finally obtain the error in $e$ and $\tau$.}
\\

\begin{figure*}
    \centering
    \begin{tabular}{cc}
        (a) Lennard-Jones (100) plane & (b) Lennard-Jones (111) plane \\
        \includegraphics[width=0.5\linewidth]{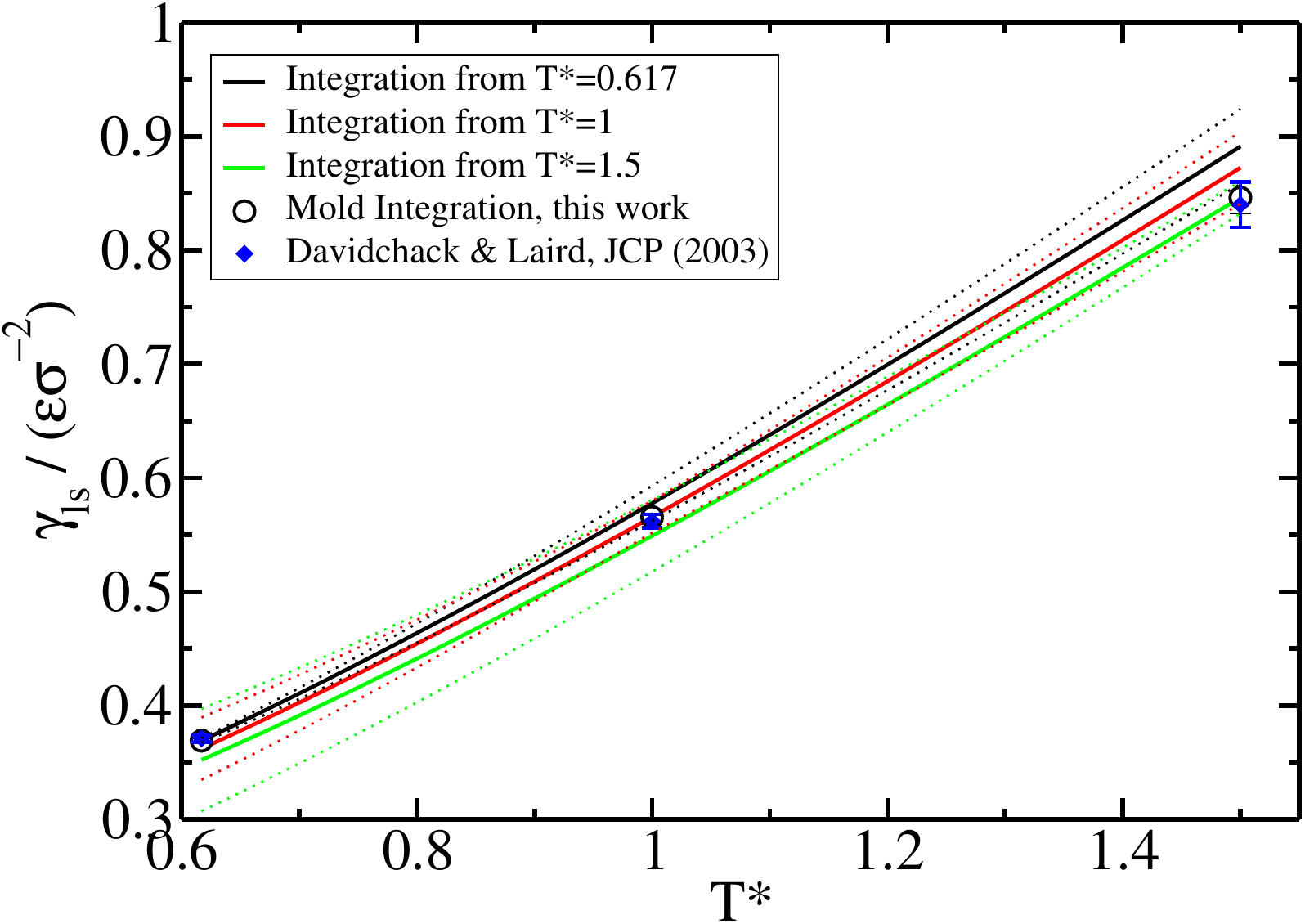} & \includegraphics[width=0.5\linewidth]{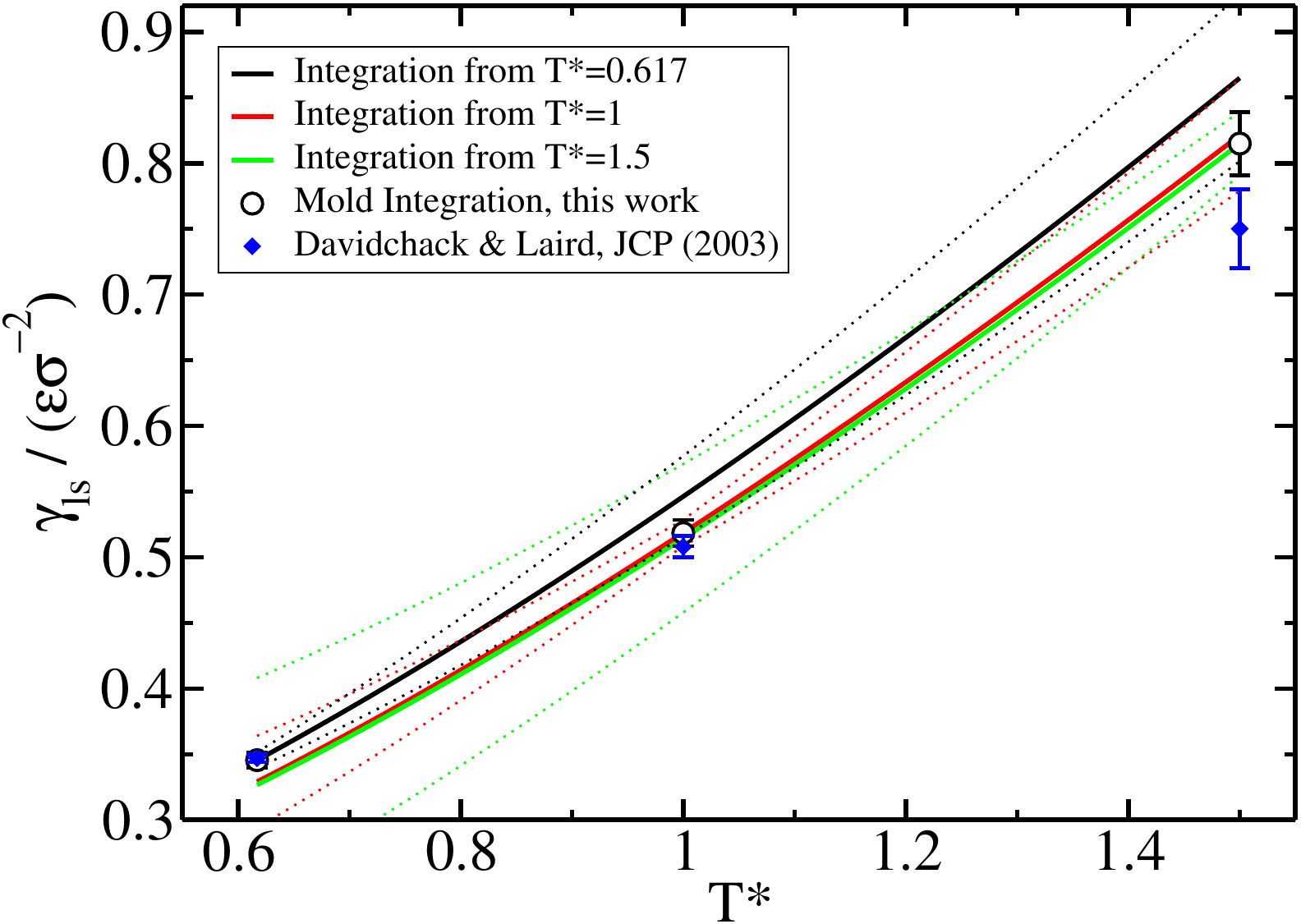}
    \end{tabular}
    \caption{Predicted interfacial free energy as a function of $T^*$ along the coexistence line for (a) the (100) Lennard-Jones plane and (b) the (111) Lennard-Jones plane of the fcc crystal phase. The continuous lines indicate the predictions obtained from Eq. \ref{integration} using the Mold Integration calculations as integration limits, while the dashed lines depict the uncertainty limits for each of the estimations from different temperatures. We also plot the direct calculations of $\gamma$ from Ref. \cite{davidchack2003direct} using the Cleaving technique. Note that for both planes, the Mold Integration $\gamma_{ls}$ and those from Ref. \cite{davidchack2003direct} overlap, \textcolor{black}{excepting the one at highest T$^*$ for the (111) plane}.}
    \label{fig2}
\end{figure*}

Next, we apply this methodology to the (100) and (111) fcc crystal planes. In Figure \ref{fig2} we show the predicted $\gamma_{ls}$ from thermodynamic integration along with direct calculations of $\gamma_{ls}$ using the Cleaving method \cite{davidchack2003direct} and Mold Integration (this work). We once again obtain a good prediction of the interfacial free energy (\textcolor{black}{the predictions from Gibbs-Cahn integration match with direct calculations within the error limits}), this time using our own estimates of $\gamma_{ls}$ as reference values for the integration (see Supplementary Material Section SIII for details on our calculations of $\gamma_{ls}$ through the Mold Integration technique). Similarly to Ref. \cite{laird2009determination}, the results for the (111) plane have a greater deviation from the calculated $\gamma_{ls}$ than those for the (110) and (100) planes: the black curve in Figure \ref{fig1}(d) 
for the (110) plane deviates 0.03$\epsilon/\sigma^2$ and the black curve in Figure \ref{fig2}(b) deviates 0.05$\epsilon/\sigma^2$ for the (111) one. These greater deviations of $\gamma_{ls}$ only appear when the target temperature changes by a factor of 2.5 the reference temperature from which the integration is performed. In this aforementioned example, the prediction is evaluated at $T^*=1.5$ when the integration begins at $T^*=0.617$
.
\\

\begin{figure*}
    \centering
    \begin{tabular}{cccc}
        (a) & (b) & (c) & (d) \\
        \includegraphics[width=0.24\linewidth]{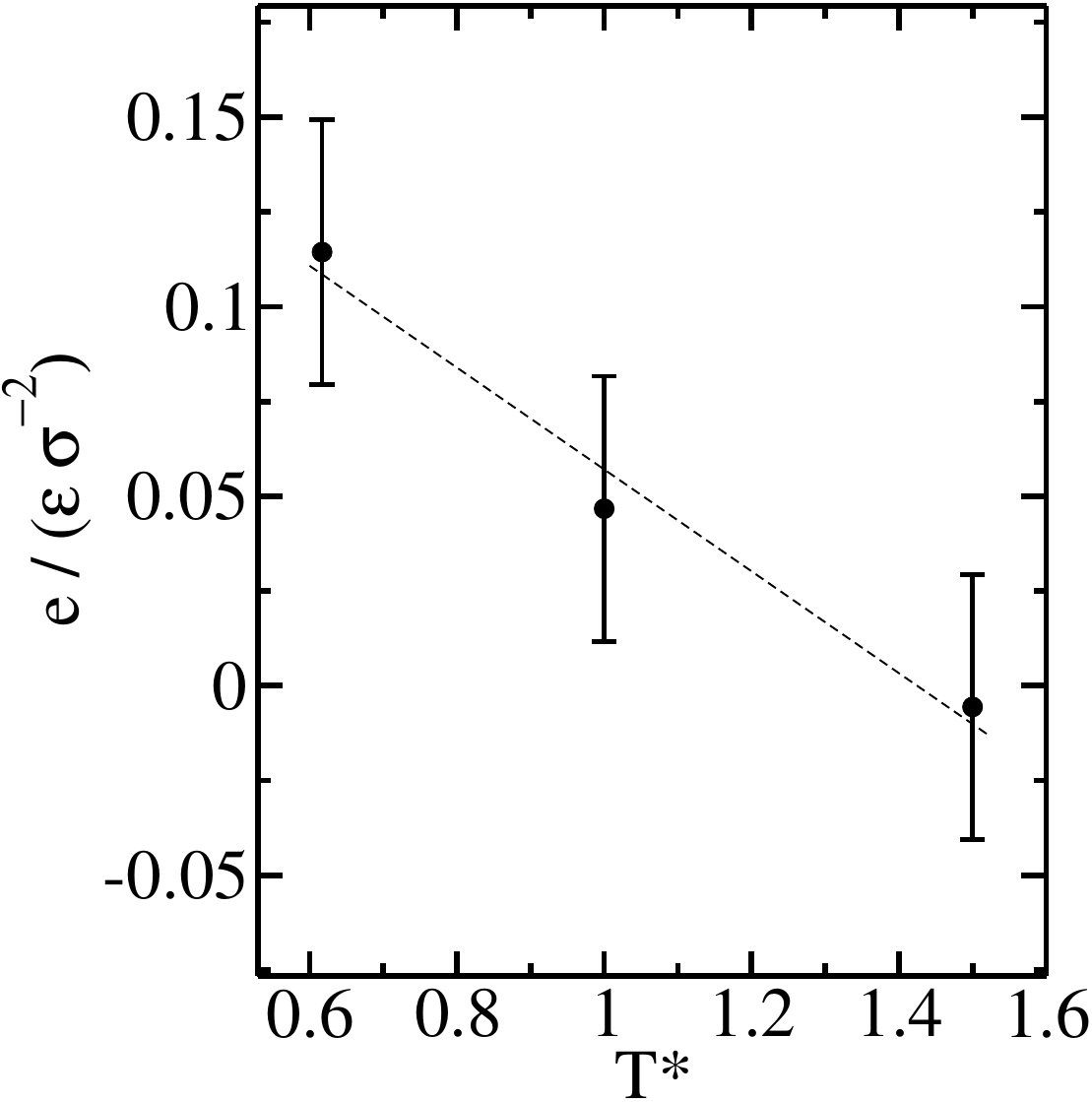} & \includegraphics[width=0.24\linewidth]{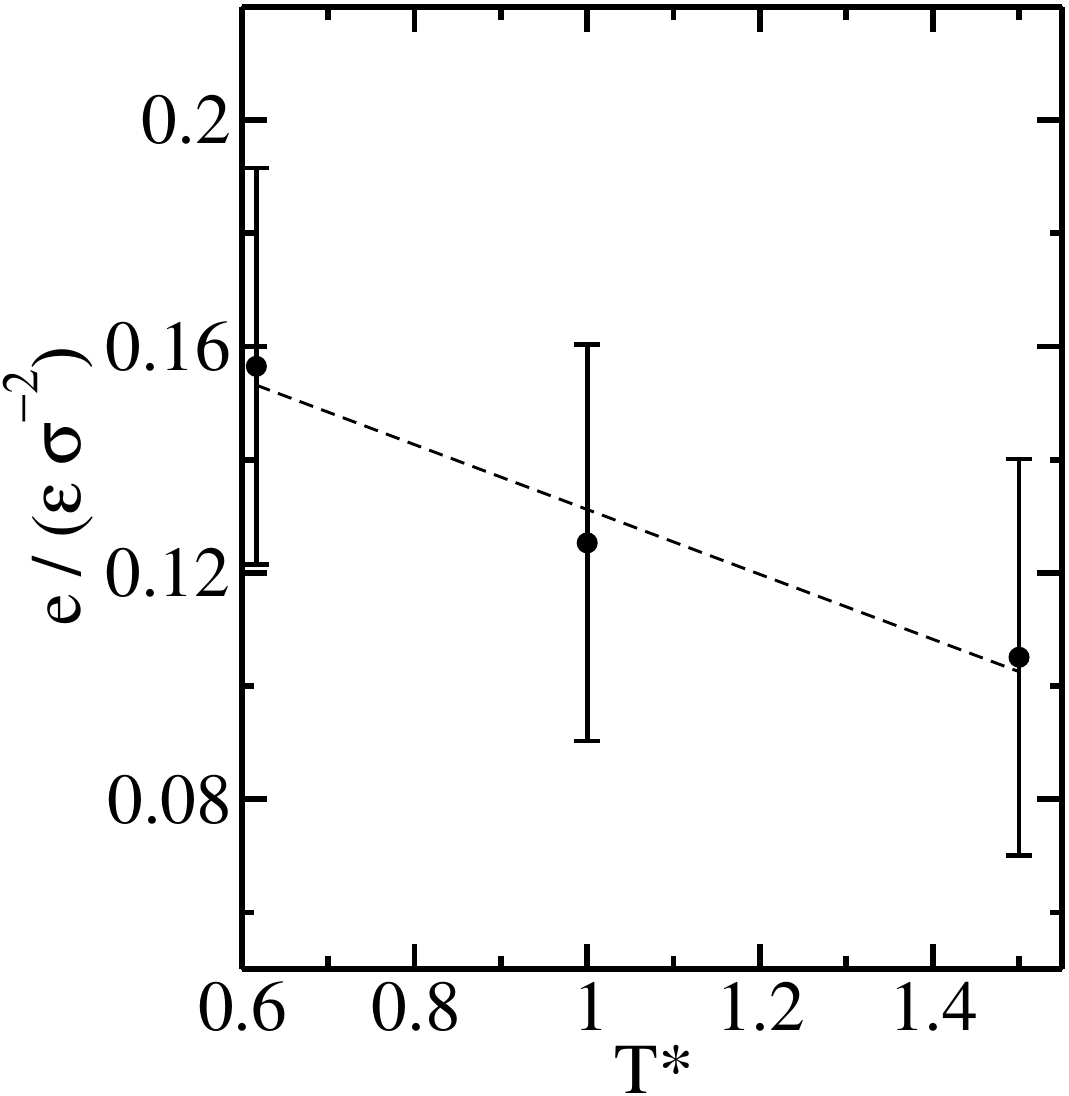} & \includegraphics[width=0.24\linewidth]{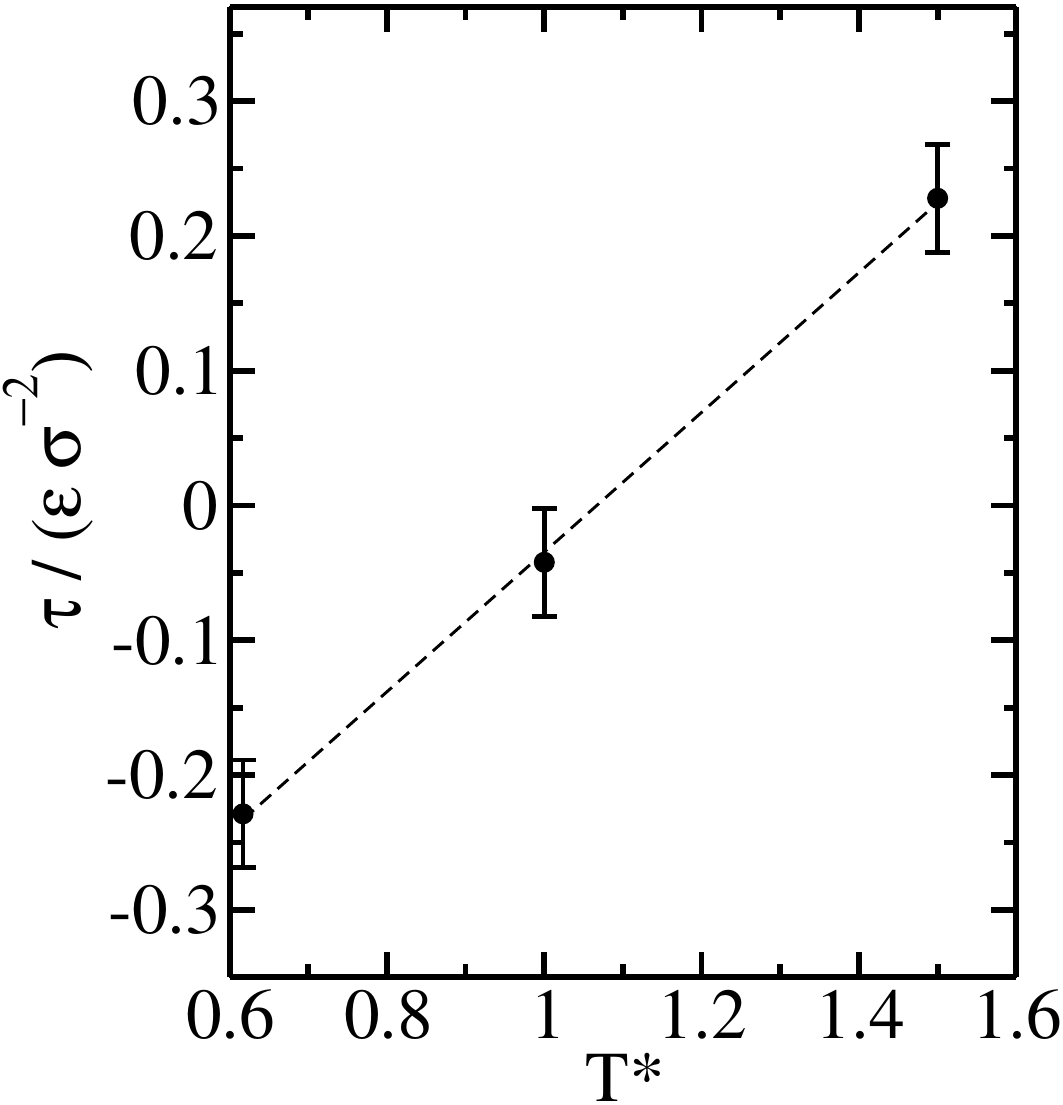} & \includegraphics[width=0.24\linewidth]{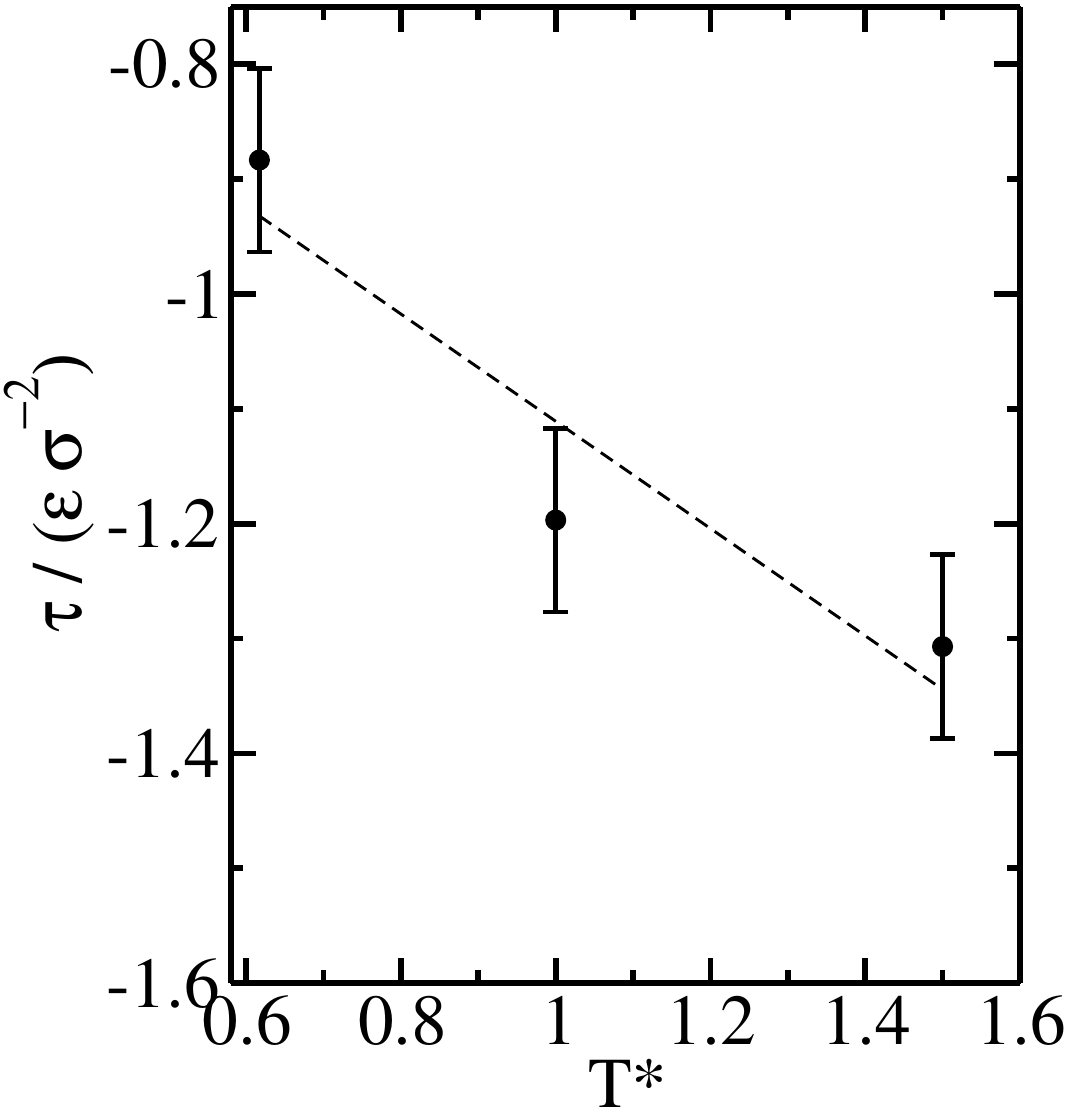}
    \end{tabular}
    \caption{Surface excess energy as a function of temperature for the Lennard-Jones liquid-solid interface of the (100) (a) and (111) (b) planes. Interfacial stress as a function of temperature for the Lennard-Jones liquid-solid interface of the (100) (c) and (111) (d) planes.}
    \label{figeandtaulj}
\end{figure*}

Additionally, we show all the individual fits for $e$ and $\tau$ as a function of $T^*$ for the (110) and (111) planes plotted in Figure \ref{figeandtaulj}. Furthermore, we provide the data for $e$ at each temperature for the three studied planes in Table \ref{tab_eandtaudata_lj}. As can be attained from such $e$ and $\tau$ data, all the values follow a monotonic trend along the coexistence line. It is remarkable how the interfacial stress is much lower for the case of the (100) plane (and even goes from negative to positive values, Figure \ref{figeandtaulj}(c)), while it is not negligible for the \textcolor{black}{(110)} and (111). This reduces the uncertainty in the integration, since the second term in the right hand side of Eq. \ref{integration} contributes very small values to the prediction of $\gamma_{ls}$ along the coexistence line. \textcolor{black}{It is also notable how the difference in the uncertainty obtained for $e$ and $\tau$ obtained from DC simulations differs between our calculations and those from Ref. \cite{laird2009determination}, which are plotted along our results in Figure \ref{fig1}. In our calculations we directly estimate the error bar from the standard deviation along the simulation. This discrepancy may arise from the fact that the system sizes used in Ref. \cite{laird2009determination} are about an order of magnitude greater, as well as the possibility of performing even more computationally intensive simulations, although the length of the simulations was not specified in Ref. \cite{laird2009determination}. Moreover, we provide direct comparison of our results, with those from Ref. \cite{laird2009determination}. To perform Gibbs-Cahn integration with such data, we fitted the $e$ and $\tau$ values provided in Ref. \cite{laird2009determination} with second order polynomial fits, as the trend of that data suggests. In Figures S5, S6 and S7 we probe how the difference between our results and those from Ref. \cite{laird2009determination} is minimal, and our results are only quantitatively less accurate for the (111) plane, despite the overall larger uncertainty in the determination in $e$ and $\tau$.}
\\
\begin{table}[h]
    \centering
    \begin{tabular}{m{1.7cm}|c|c|c|c}
         & $T^*$ & (100) & (110) & (111) \\ \hline
        \multirow{3}{4em}{\mbox{$e$ / ($\epsilon$ $\sigma^{-2}$)}} & 0.617 & 0.11(3) & 0.16(4) & 0.16(4) \\
         & 1 & 0.05(3) & 0.16(4) & 0.13(4) \\
         & 1.5 & -0.01(3) & 0.15(4) & 0.11(4) \\ \hline
        \multirow{3}{4em}{\mbox{$\tau$ / ($\epsilon$ $\sigma^{-2}$)}} & 0.617  & -0.23(4) & -0.90(7) & -0.88(8) \\
         & 1 & -0.04(4) & -0.92(7) & -1.20(8) \\
         & 1.5 & 0.23(4) & -1.17(7) & -1.31(8)
    \end{tabular}
    \caption{$e$ and $\tau$ data for all of the different states considered in this study for the Lennard-Jones system.}
    \label{tab_eandtaudata_lj}
\end{table}

We independently estimate the interfacial free energy of the three different states studied here ($T^*=0.617,\ T^*=1,\ T^*=1.5$), which are the same as those evaluated in Ref. \cite{davidchack2003direct}. For this purpose, we perform Mold Integration (MI) calculations (see Supplementary Material, section SIII for further details on this technique and calculations). We plot our $\gamma_{ls}$ values from Mold Integration (black circles filled in white) along with those from Ref. \cite{davidchack2003direct} (blue diamonds) in Figure \ref{fig2}. Overall, we find excellent agreement between our calculations and those from Ref. \cite{davidchack2003direct} for all the cases except for the (111) plane at $T^*=1.5$, for which we obtain a moderately higher value. 
Interestingly, the value obtained from Mold Integration is in better agreement with our estimations from the Gibbs-Cahn thermodynamic integration than with the Cleaving values from Ref. \cite{davidchack2003direct}, especially with the fit obtained from $T^*=1$ (red curve in Figure \ref{fig2}(b)). It must be noted that, unlike for the case of the (110) plane (where $\gamma_{ls}$ values were taken from Ref. \cite{davidchack2003direct}), for the (100) and (111) we use MI \textcolor{black}{values as integration starting points ($\gamma_0$, and $T_0$) to} solve Eq. \ref{integration}.
\\

The predictions from thermodynamic integration yield results \textcolor{black}{that overlap with those directly evaluated within the error bars} (Figures \ref{fig1}(d) and \ref{fig2}), however the deviation becomes larger for values estimated from very distant temperatures. We find that in the Lennard-Jones system, all the $\gamma_{ls}$ thermodynamic integration predictions match with the directly measured values for the 3 studied planes (Figures \ref{fig1}(d) and \ref{fig2}).

\subsection{Thermodynamic integration of $\gamma$ for the mW water model along its melting line}

\begin{figure}
    \centering
    \includegraphics[width=\linewidth]{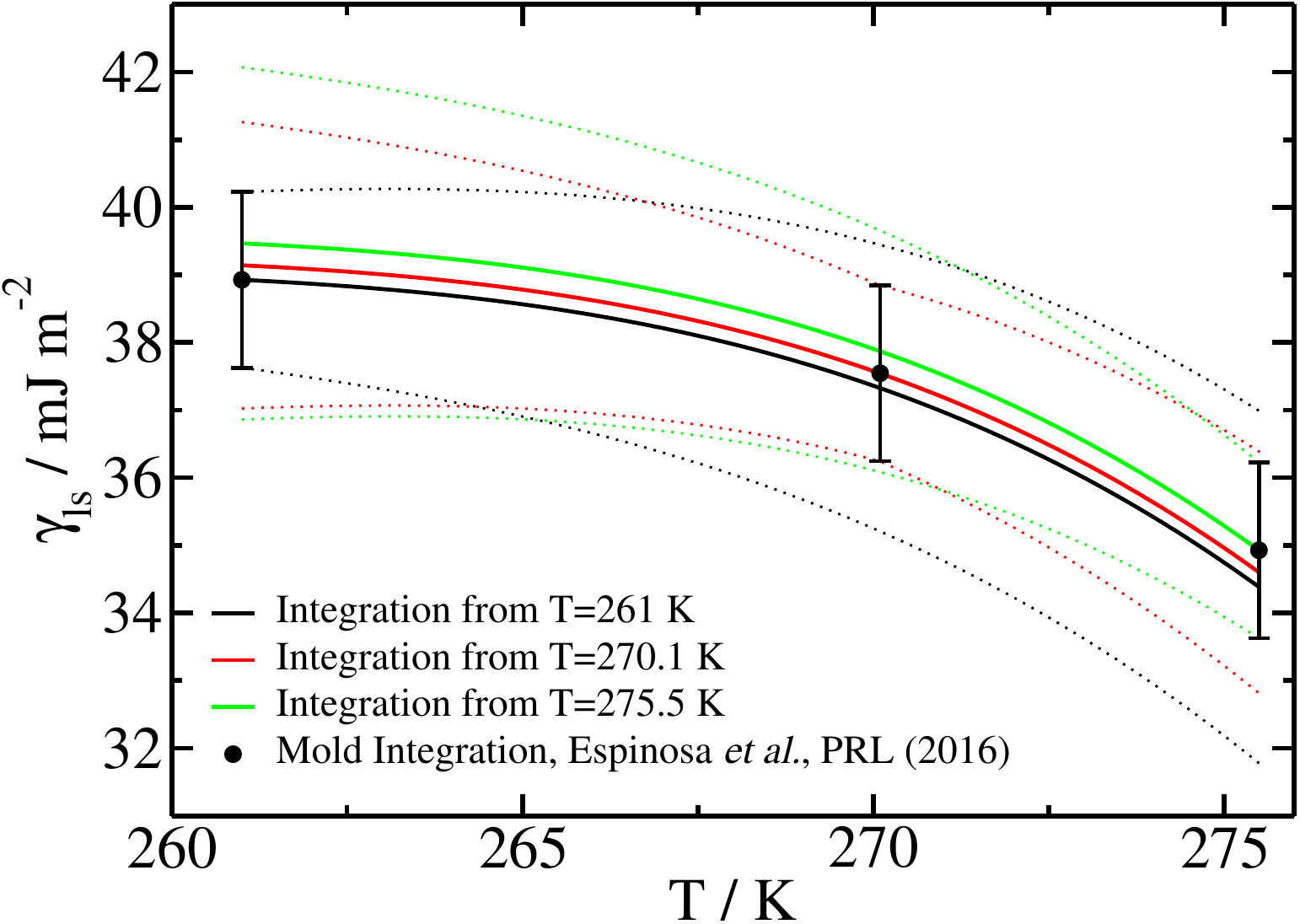}
    \caption{Predicted interfacial free energy against temperature (along the melting line) from Gibbs-Cahn integration (Eq. \ref{integration}) for the ice Ih basal plane of the mW water model. Direct calculations of $\gamma_{ls}$ using Mold Integration \cite{espinosa2016interfacial} (black circles) have been been taken as integration limits.}
    \label{fig3}
\end{figure}

We now proceed to test the Gibbs-Cahn integration methodology along the liquid-solid coexistence line of a more complex case, the mW coarse-grained water model. This monoatomic model reproduces the behaviour of water because the tetrahedral coordination of the atoms is favored by adding a three-body term that penalizes non-tetrahedral coordinations (see Supplementary Material SI). We aim to predict $\gamma_{ls}$ for the ice Ih basal (0001) plane, for which previous calculations from Espinosa \textit{et al.} \cite{espinosa2016interfacial} using Mold Integration have already established the dependence of $\gamma_{ls}$ with $T$ along the melting line. Analogously to the BG Lennard-Jones case, we perform independent NVT DC and NpT bulk simulations in order to obtain the dependence of $e$, $\tau$, and $\rho_s$ with temperature, and later solve Eq. \ref{integration}, using the values from Ref. \cite{espinosa2016interfacial} as \textcolor{black}{starting points for the integration ($\gamma_{ls,0}$ and $T_0$ in equation \ref{integration}).}
\\

\begin{figure}
    \centering
    (a) \\
    \includegraphics[width=\linewidth]{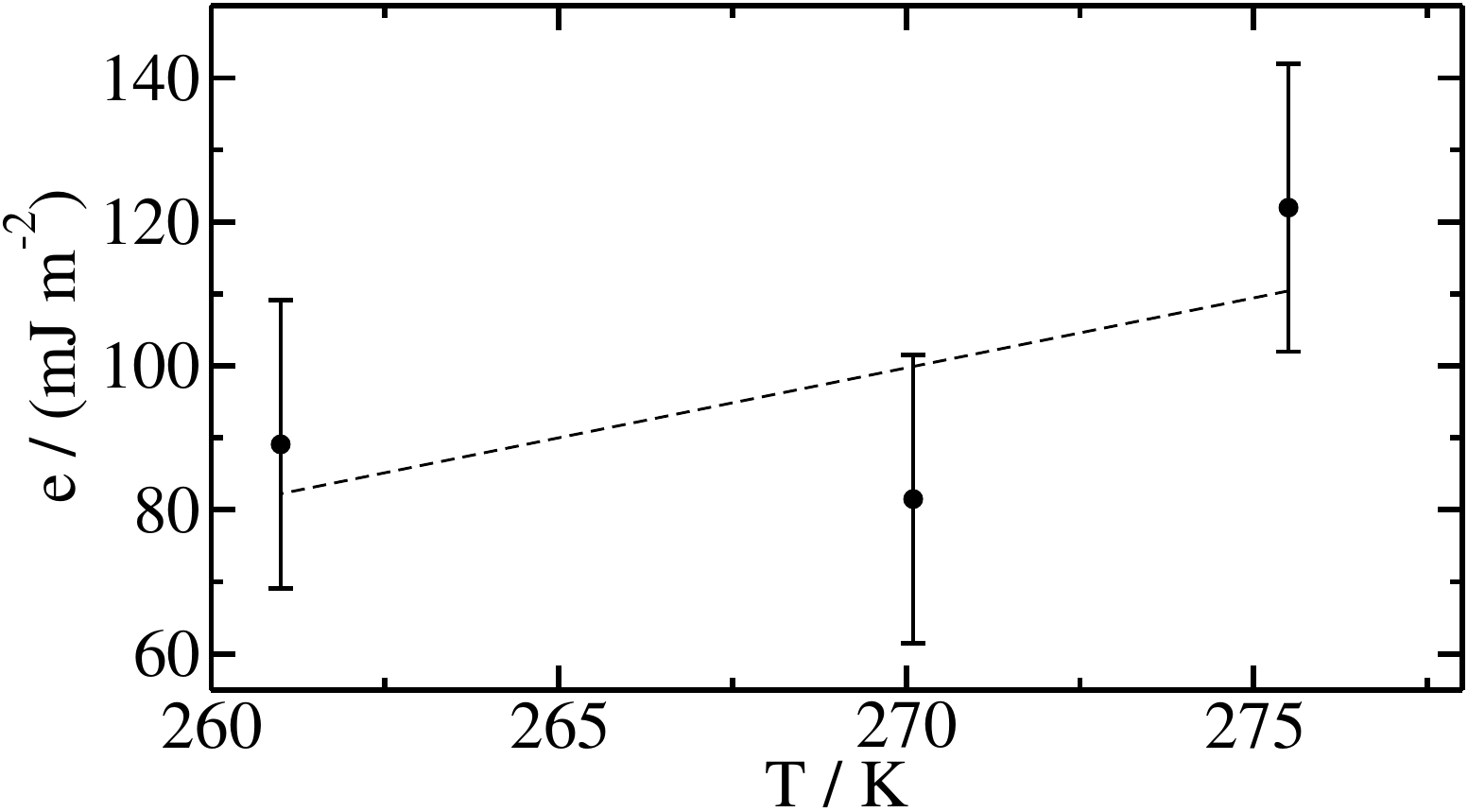} \\
    (b) \\
    \includegraphics[width=\linewidth]{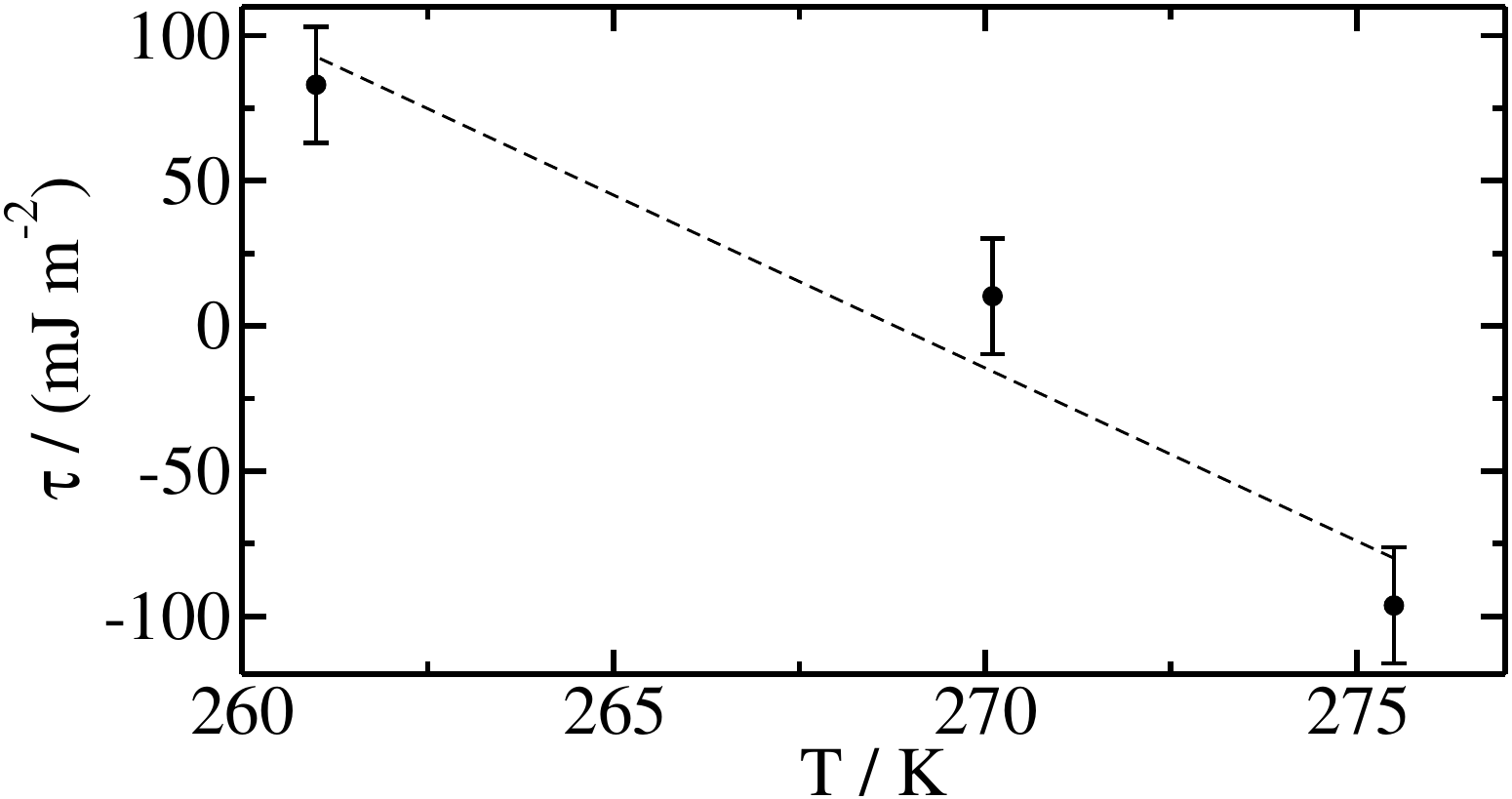}
    \caption{(a) Surface excess energy ($e$) against $T$ for the ice Ih-fluid of the mW water model with the crystal phase exposing the basal plane to the liquid. (b) Interfacial stress ($\tau$) against $T$ for the ice Ih-fluid of the mW water model with the crystal phase exposing the basal plane to the liquid.}
    \label{fig_etau_mw}
\end{figure}

\begin{table}[h!]
    \centering
    \begin{tabular}{c|c|c|c|c}
         & $T\ /\ K$ & $P\ /\ bar$ & $e$ / (mJ m$^{-2}$) & $\tau$ / (mJ m$^{-2}$) \\ \hline
        \multirow{3}{4em}{Ice Ih mW} & 275.5 & 1 & 122(20) & -96(20) \\
         & 270.1 & 2000 & 81(20) & 10(20)  \\
         & 261 & 5000 & 89(20) & 83(20) 
    \end{tabular}
    \caption{$e$ and $\tau$ values for all of the different states considered in this study for the ice Ih basal plane of the mW water model.}
    \label{tab_eandtaudata_mw}
\end{table}

In Figure \ref{fig3} we show the results of $\gamma_{ls}$ from thermodynamic integration for the mW model. We observe that the Gibbs-Cahn integration is able to describe the change of interfacial free energy with temperature along the coexistence line. \textcolor{black}{We find that the curvature of $\gamma_{ls}$} with T deviates from a linear trend in a more pronounced way than for the BG Lennard-Jones system (see Figures \ref{fig1}(d) and \ref{fig2}), and the value of $\gamma_{ls}$ decreases with T (for the temperature range studied here). The fact that the Gibbs-Cahn integration predicts this behaviour proves the power of this approach to quantify the dependence of the interfacial free energy along a coexistence line. This is of great relevance for the case of solid-liquid interfaces, where $\gamma_{ls}$ calculations are challenging and it is required the implementation of advanced simulation techniques
\cite{broughton1986molecular,fernandez2012equilibrium,angioletti2010solid,espinosa2014mold,tejedor2024mold,hoyt2001method,benjamin2014crystal,bultmann2020computation}. 
For the mW model, the $e$ and $\tau$ fits are shown in Figure \ref{fig_etau_mw}, and the corresponding data can also be found in Table \ref{tab_eandtaudata_mw}. From Figure \ref{fig_etau_mw} can be seen that, differently to the case of the Lennard-Jones system, $e$ increases with temperature. The interplay between increasing $e$ and decreasing $\tau$ along the coexistence line for this temperature range yields the singular trend of the interfacial free energy that, apart from direct measurements, can be recovered using the Gibbs-Cahn integration method.

\subsection{Thermodynamic integration of $\gamma$ along the liquid-vapor coexistence line}

Estimating the liquid-vapor interfacial free energy ($\gamma_{lv}$) along the coexistence line is considerably simpler than it is for the liquid-solid interface. By means of a DC simulation containing the two coexisting phases in the NVT ensemble, $\gamma_{lv}$ can be determined through the relation 

\begin{equation}
        \gamma_{lv} = \frac{L_x}{2}(\overline{p}_N-\overline{p}_T).
        \label{kirkwood}
\end{equation}

Nevertheless, as a proof of concept, we apply here the Gibbs-Cahn thermodynamic integration method to show how $\gamma_{lv}$ can be also easily determined along the coexistence line with a similar approach as that shown for liquid-solid interfaces. For this goal, one needs to compute the excess interfacial energy along temperature at coexisting conditions (Eq. \ref{integrallv}). We then proceed by first performing liquid-vapor DC simulations in the canonical ensemble, from which we obtain $E$, as well as the coexistence pressure at the fixed temperature. Once we know the coexistence pressure, we perform NpT simulations of the bulk liquid and vapor phases under such conditions separately, in order to obtain $\rho^E$ for both phases, \textcolor{black}{which with the use of Eq. \ref{excess_energy} we obtain $e$. We can then} fit $e$ as a function of the inverse of temperature, which will be integrated (Eq. \ref{integrallv}). In this case, the integration limit conditions ($\gamma_0$ at $T_0$ in Eq. \ref{integrallv}) are also extracted from NVT DC simulations applying the Kirkwood-Buff equation (Eq. \ref{kirkwood}). This equation is equivalent to Eq. \ref{taueq}, used in liquid-solid equilibrium to compute the excess interfacial stress \cite{di2020shuttleworth}.
\\

\begin{figure}
    \centering
    (a) Lennard-Jones liquid-vapor \\
    \includegraphics[width=\linewidth]{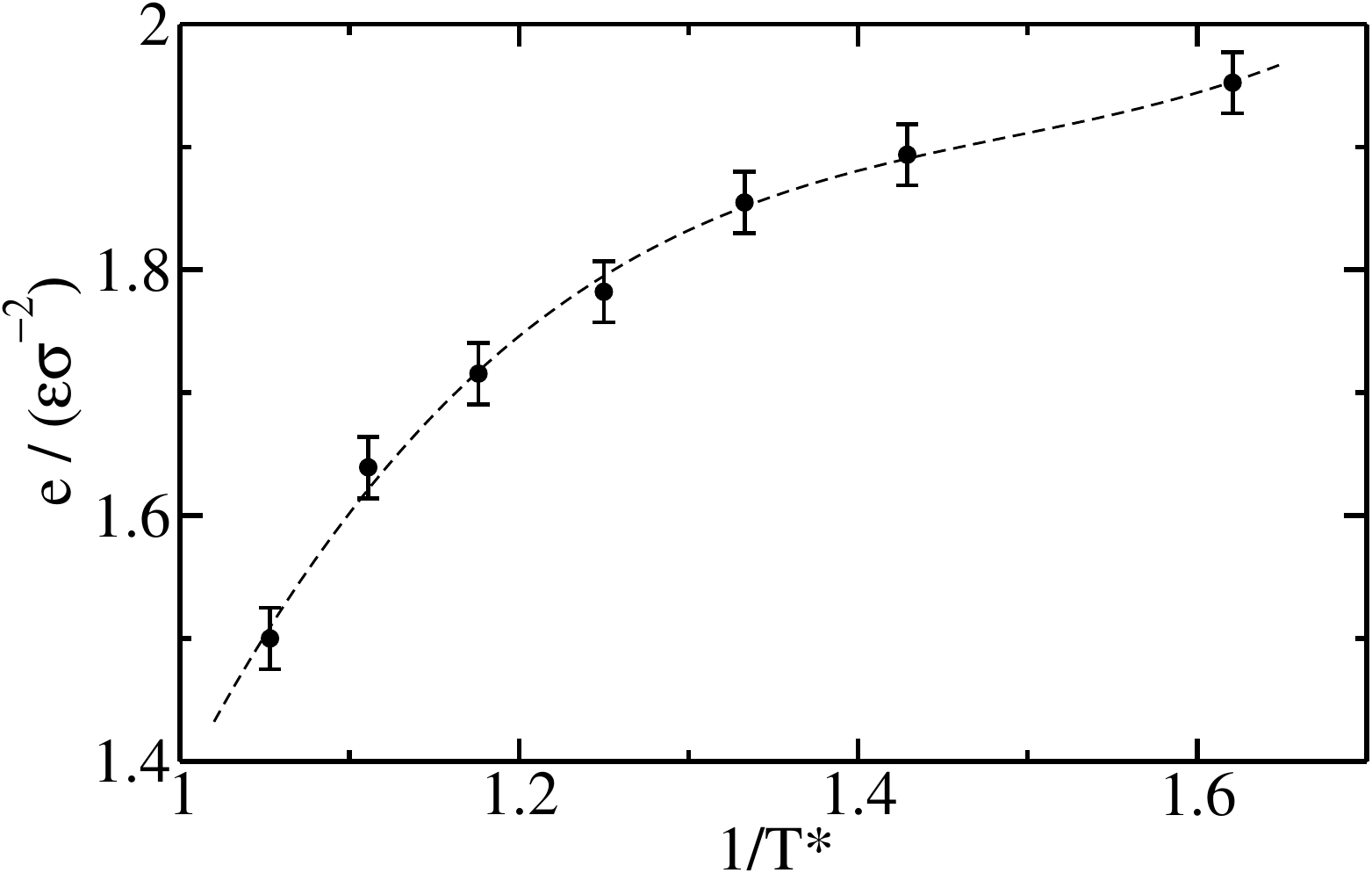} \\
    (b) TIP4P/2005 liquid-vapor \\
    \includegraphics[width=\linewidth]{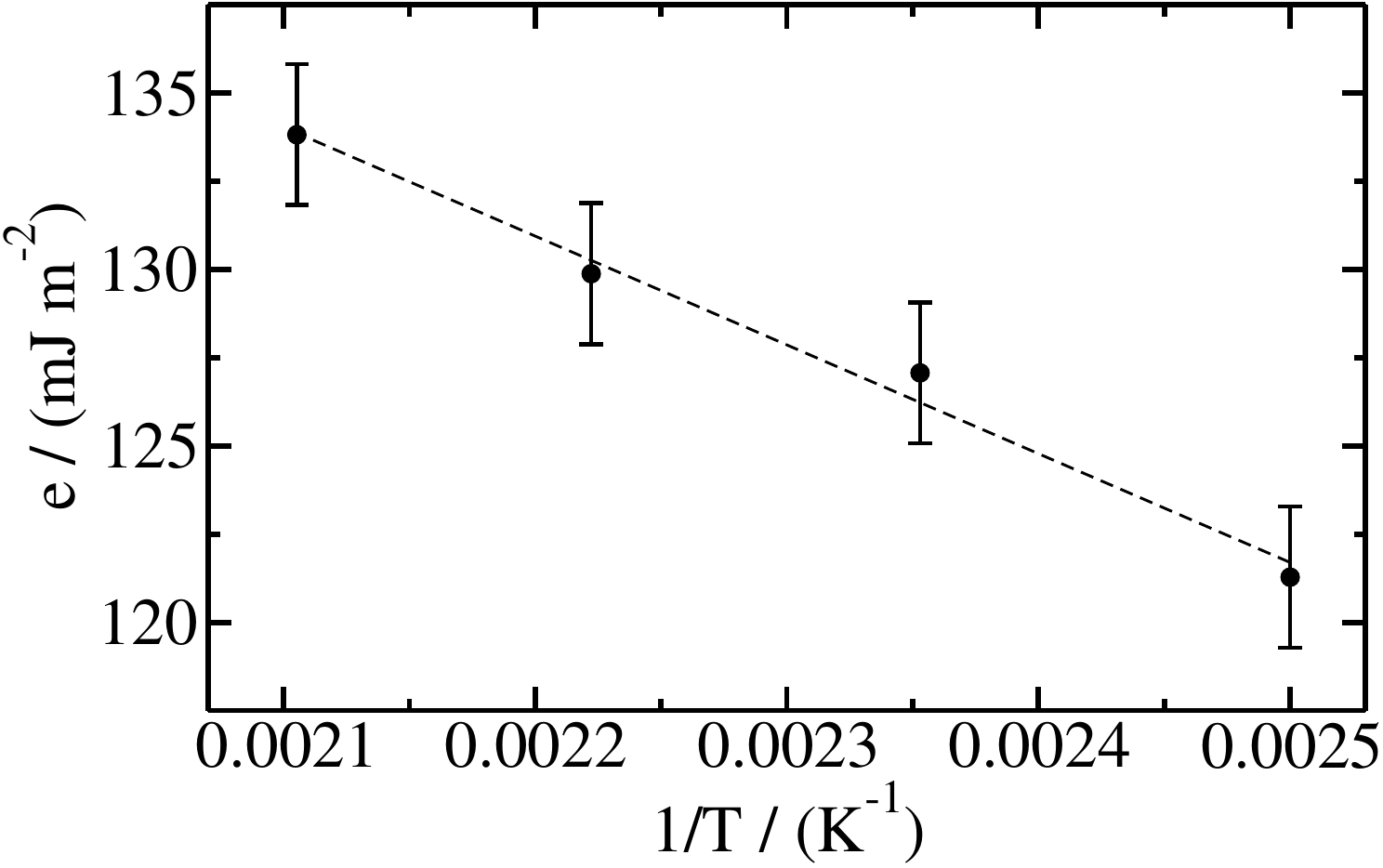}
    \caption{(a) Surface excess energy ($e$) against inverse of temperature (at the corresponding coexisting pressure for each temperature) for the Lennard-Jones liquid-vapor interface. Filled points are direct estimations while the dashed line indicates the fit to the data. \textcolor{black}{In this case, to better follow the trend, a 3rd order polynomial fit was used.} (b) $e$ against inverse of temperature for the TIP4P/2005 water model liquid-vapor interface. Filled points are direct estimations of $e$ while the dashed line depicts the linear fit to the obtained data.}
    \label{ev1t_lj_tip4p}
\end{figure}

\begin{figure}
    \centering
    (a) Lennard-Jones liquid-vapor \\
    \includegraphics[width=\linewidth]{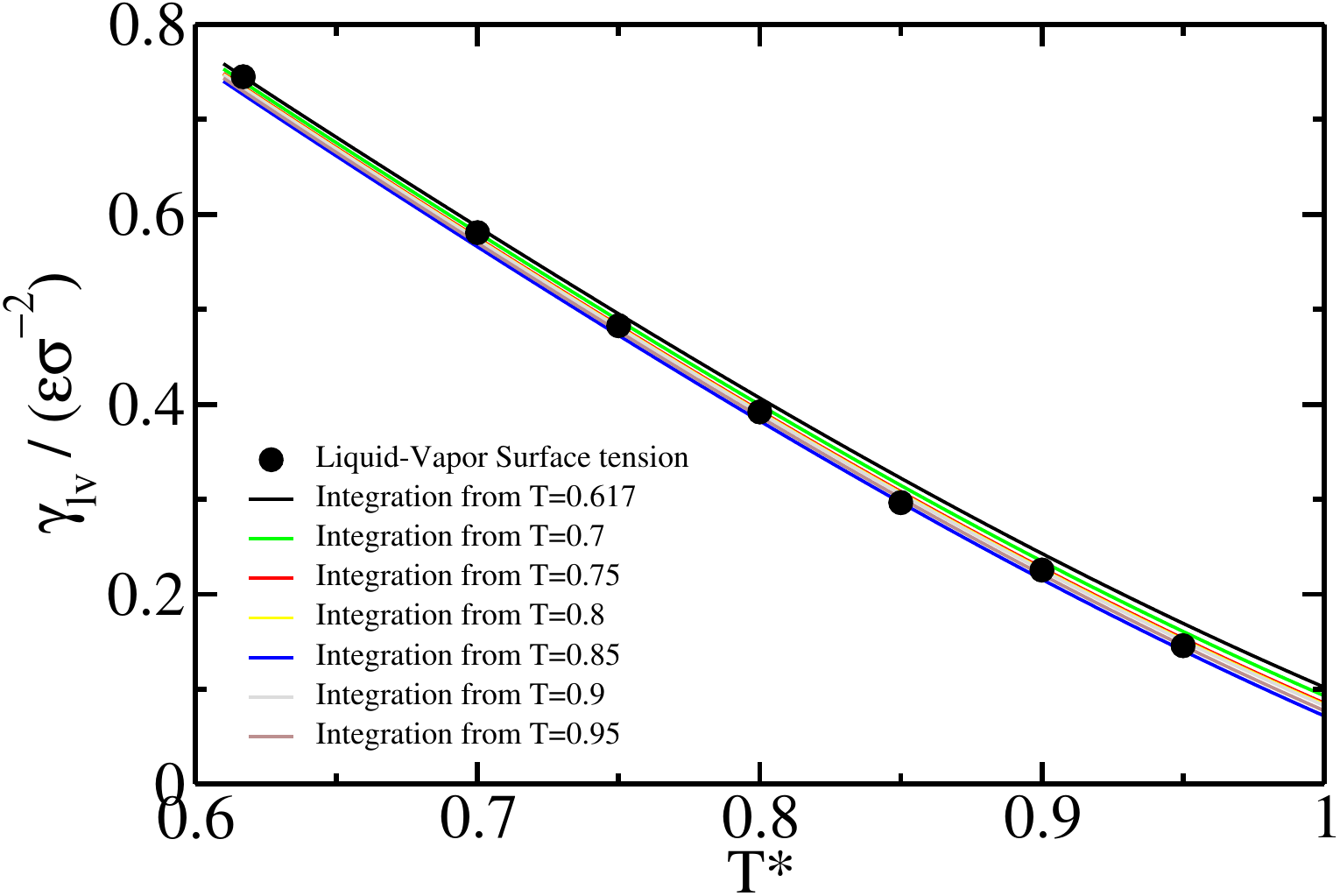} \\
    (b) TIP4P/2005 liquid-vapor \\
    \includegraphics[width=\linewidth]{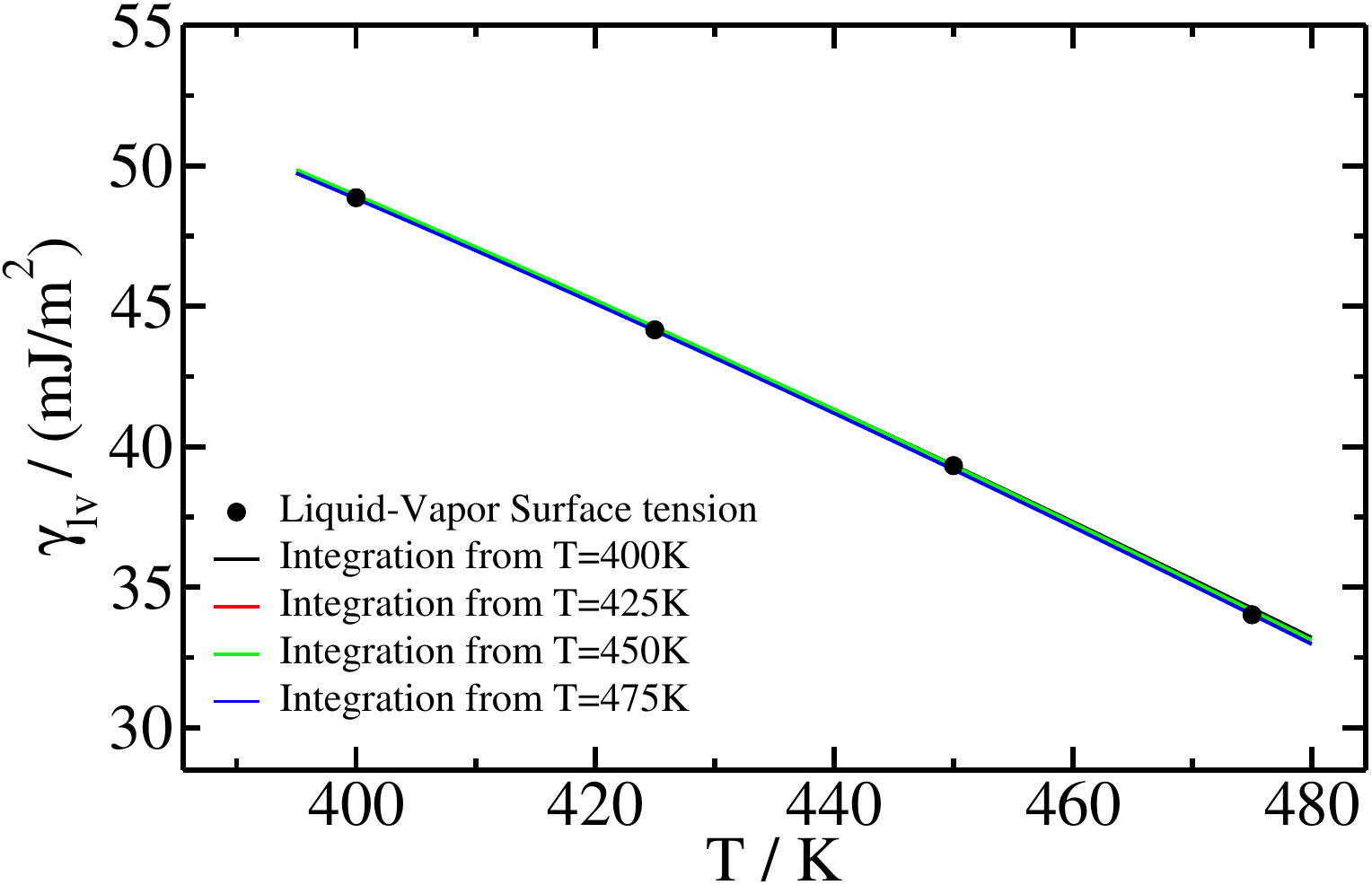}
    \caption{(a) Interfacial free energy against temperature along the BG Lennard-Jones liquid-vapor coexistence line as predicted through thermodynamic integration (Eq. \ref{integrallv}) from the states indicated in the legend. Direct calculations of $\gamma_{lv}$ through Eq. \ref{kirkwood} are also included as black circles. (b) Interfacial free energy against temperature along the TIP4P/2005 water model liquid-vapor coexistence line. Thermodynamic integration estimates are represented by curves whereas direct calculations of $\gamma_{lv}$ by black circles.}
    \label{fig4}
\end{figure}

\begin{table}[]
    \centering
    \begin{tabular}{c|c}
        $T^*$ & $e$ / ($\epsilon\ \sigma^{-2}$)  \\ \hline
        0.617 & 1.952(25) \\
        0.7 & 1.894(25) \\
        0.75 & 1.855(25) \\
        0.8	& 1.782(25) \\
        0.85 & 1.715(25) \\
        0.9	& 1.639(25) \\
        0.95 & 1.500(25) 
    \end{tabular}
    \caption{Surface excess energy ($e$) values for the Lennard-Jones liquid-vapor interface at the different temperatures studied.}
    \label{tab_edata_lj}
\end{table}

\begin{table}[]
    \centering
    \begin{tabular}{c|c}
        $T\ /\ K$ & $e$ / ($mJ\ m^{-2}$)  \\ \hline
        400 & 121(2) \\
        425 & 127(2)\\
        450 & 130(2) \\
        475	& 134(2)  
    \end{tabular}
    \caption{Surface excess energy ($e$) values for the TIP4P/2005 liquid-vapor interface at the different temperatures studied.}
    \label{tab_edata_tip4p}
\end{table}

We perform the aforementioned simulations for the BG Lennard-Jones system and the TIP4P/2005 water model, and then make use of thermodynamic integration (Eq. \ref{integrallv}) to predict $\gamma_{lv}$ along the coexistence line. We show the employed $e$ against 1/T fits in Figure \ref{ev1t_lj_tip4p}, as well as include the $e$ values for each temperature in Tables \ref{tab_edata_lj} (for the Lennard-Jones model) and \ref{tab_edata_tip4p} (for the TIP4P/2005 model). Our $e$ calculations for the TIP4P/2005 model are in good agreement with the those from Hrahsheh \textit{et al.} \cite{hrahsheh2024second}, \textcolor{black}{since we perfectly capure the trend obtained from single-state calculations}. In Figure \ref{fig4}(a) we show how the integration method for the BG Lennard-Jones potential yields very accurate predictions of $\gamma_{lv}$ with temperature, independently of the starting integration temperature. We also show that for the TIP4P/2005 (Figure \ref{fig4}(b)), the prediction of $\gamma_{lv}$ from the different states also provides accurate interfacial free energies along the coexistence line.
These results show that only the interfacial energy, $e$, is needed to determine the change in $\gamma_{lv}$ across liquid-vapor coexistence lines (as shown by Eq. \ref{integrallv}
We also clarify that, since we are using a truncated TIP4P/2005 potential, we underestimate the critical point and, consequently, the interfacial free energy values \cite{vega2007surface,wang2019second,gorfer2023high} that we report in this work. Nonetheless, our calculations are consistent since all our simulations are performed under the same approximations, and the direct and indirect estimates of $\gamma_{lv}$ reach excellent agreement.
\\

It must be noted here that the application of the thermodynamic integration scheme to predict liquid-vapor interfacial free energies along the coexistence line is not an efficient method, since in order to provide a prediction of $\gamma_{lv}$ one must perform NVT DC simulations at the states that will be later inferred. However, by doing these simulations, $\gamma_{lv}$ can already be obtained from Eq. \ref{kirkwood}, making the subsequent integration calculations irrelevant. Nevertheless, these calculations show how this framework is also valid for liquid-vapor interfaces.

\subsection{Evaluation of thermodynamic integration equations}

\begin{figure}
    \centering
    \includegraphics[width=\linewidth]{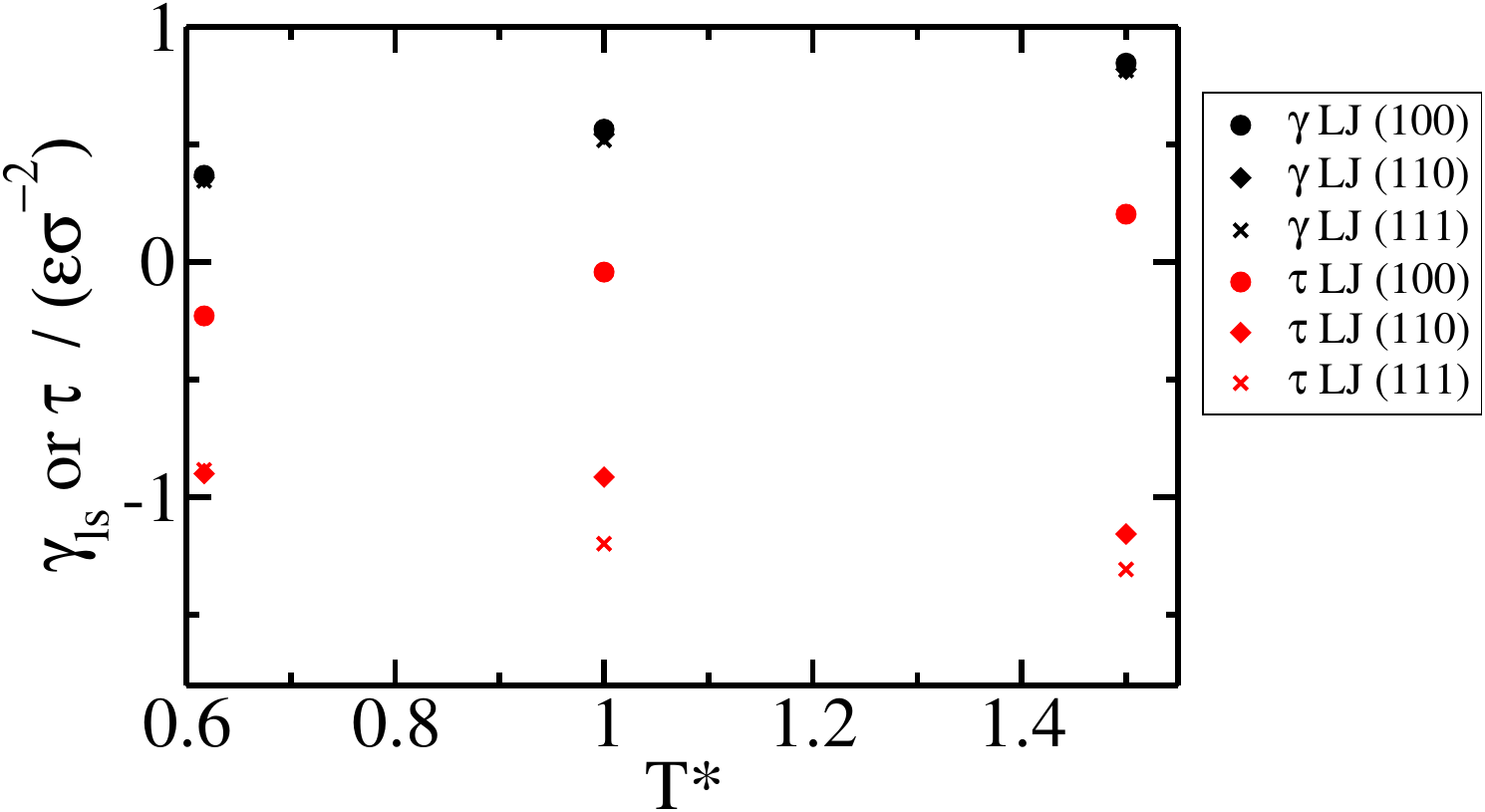}
    \caption{Liquid-solid interfacial free energy ($\gamma_{ls}$) and interfacial stress ($\tau$) as a function of temperature for the different orientations studied (as indicated in the legend) with the Lennard-Jones BG potential. The interfacial free energy values are calculated by us through Mold Integration calculations, excepting the (110) orientation, which is taken from Ref. \cite{davidchack2003direct}.}
    \label{fig5}
\end{figure}

As it has been previously argued \cite{di2020shuttleworth}, the expressions, methods, and complexity to estimate the interfacial free energy for liquid-vapor and liquid-solid interfaces are completely different. For the case of liquid-vapor interfaces, the calculation is straightforward and can be carried out by using Eq. \ref{kirkwood}. However we note that the interfacial free energy in liquid-solid interfaces cannot be calculated through such relation \cite{di2020shuttleworth} rather than through specialised simulations techniques \cite{broughton1986molecular,fernandez2012equilibrium,angioletti2010solid,espinosa2014mold,tejedor2024mold,hoyt2001method,benjamin2014crystal,bultmann2020computation}. Here, we illustrate how $\gamma_{ls}$ should not be obtained through Eq. \ref{kirkwood}. In Figure \ref{fig5} we plot both $\gamma_{ls}$ and $\tau$ along the liquid-solid coexistence line plotted as a function of temperature for the different crystal planes studied with the BG Lennard-Jones potential. It is evident that the two quantities do not match. In fact, the interfacial stress takes negative values for most of the states and orientations we examined, which is unphysical ($\gamma_{ls}$ must be positive, otherwise demixing into two phases would not occur).
\\

\textcolor{black}{Figure 9 highlights the importance of employing an adequate thermodynamic derivation or simulation method to attain reliable interfacial free energies for a given system. 
Since this is a fundamental conceptual error, the community should not use Eq. \ref{kirkwood} to determine the interfacial free energy between a fluid and a solid \cite{di2020shuttleworth,frenkel2013simulations}. As can be seen in Figure \ref{fig5}, that could even lead to negative values of the interfacial free energy which is of course an incorrect result.
Notice that although $\gamma$ changes at most by about ten per cent from one plane to another (at least for the LJ system) values of
$\tau$ change dramatically from one plane to another. The plane (hkl) dependence of $\gamma$ is small whereas the plane (hkl) dependence of $\tau$ is huge.
That means that one could find crystal orientations for which the stress ($\tau$) is almost zero (for instance the plane 100 of the LJ system shown in Figure \ref{figeandtaulj}(c)).
For these particular planes one can neglect the $\tau$ contribution in Eq. \ref{integrationold} (i.e omitting the second term of the square bracket) to elucidate the change of $\gamma$ along the coexistence curve (thus will introduce only an small error) to facilitate a rather simple physical (although approximate) interpretation of the changes of $\gamma_{ls}$ along the coexistence curve in terms of the surface excess energy $e$.
In retrospective this is probably what happened in our previous work \cite{montero2023minimum} with the basal plane of ice Ih and Eq. \ref{integrallv} to provide simple explanation of the origin of a minimum in the ice-water interfacial free energy.}
\\

\section{Conclusion}

In this work, we apply a robust thermodynamic integration approach for predicting interfacial free energies along coexistence lines, focusing on liquid-solid and liquid-vapor interfaces. The approach employs Molecular Dynamics simulations as a working tool, and it streamlines the calculation of $\gamma$ along coexistence lines by requiring just one initial calculation of this quantity at a reference state. Gibbs-Cahn thermodynamic integration stands as a valuable tool for characterizing one of the most critical quantities in phase transitions---the interfacial free energy---across a wide range of conditions, effectively bypassing the computational challenges and cost associated with traditional techniques for the liquid-solid interface  \cite{davidchack2003direct,sanchez2021fcc,bai2006calculation}.
\\

By successfully predicting the interfacial free energy for both liquid-solid (with Lennard-Jones and the mW model) and liquid-vapor (with Lennard-Jones and the TIP4P/2005 model) interfaces, we have demonstrated the method's versatility, making it suitable for a wide array of materials and conditions. We distinguish between interfaces with and without solid phases, which makes the thermodynamic route to predict $\gamma$ along a coexistence line different. The computational efficiency and accuracy of the method is evident from our results, which consistently agree with other computational techniques that provide direct estimates of $\gamma$.
\\

The potential applications of $\gamma$ predictions through thermodynamic integration are diverse, spanning multiple scientific disciplines, including materials science, chemistry, and physics \cite{skapski1956theory,lippmann2018determining,gunduz1985measurement,wilson2015solid}, and complements other thermodynamic pathways for single-state point calculations of $\gamma_{ls}$ \cite{laird2009determination,laird2010calculation,bultmann2020computation,espinosa2014mold}.

\section{Supplementary Material}

See the supplementary material for additional information on the Lennard-Jones, mW and TIP4P/2005 potentials details, as well as an explanation on the Direct Coexistence method, our Mold Integration calculation results, the Lennard-Jones and mW equations of state, a proof for Eq. \ref{lv1}, and a direct comparison between our thermodynamic integration predictions and those from Ref. \cite{laird2009determination}.

\section{Acknowledgements}

I.~S.-B. acknowledges funding from Derek Brewer scholarship of Emmanuel College and EPSRC Doctoral Training Programme studentship, number EP/T517847/1. J.~R.~E. acknowledges funding from Emmanuel College, the University of Cambridge, the Ramon y Cajal fellowship (RYC2021-030937-I), and the Spanish scientific plan and committee for research reference PID2022-136919NA-C33. This work has been performed using resources provided by the Cambridge Tier-2 system operated by the University of Cambridge Research Computing Service (http://www.hpc.cam.ac.uk) funded by EPSRC Tier-2 capital grant EP/P020259/1-CS170. This work has also been performed using resources provided by Archer2 (https://www.archer2.ac.uk/) funded by EPSRC Tier-2 capital grant EP/P020259/e829. E. Sanz and C.Vega acknowledge funding from grant 
PID2022-136919NB-C31 of the MICINN.
The authors thankfully acknowledge RES computational resources provided by Mare Nostrum 5 through the activity 2024-2-0002

\section{Author declarations}

The author authors have no conflicts to disclose.

\section{Data availability}

The data that support the findings of this study are available within the article and its supplementary material.

\bibliographystyle{ieeetr}

\clearpage

\counterwithin{equation}{section}
\counterwithin{table}{section}
\setcounter{section}{0}
\setcounter{figure}{0}
\renewcommand\thesection{S\Roman{section}}   
\renewcommand\thefigure{S\arabic{figure}} 
\renewcommand\thetable{S\arabic{table}} 
\renewcommand\theequation{S\arabic{equation}}    
\onecolumngrid
\setcounter{page}{1}
\begin{center}
    {\large \textbf{Supplementary Material: Predictions of the interfacial free energy along the coexistence line from single-state calculations} \\ \par} \vspace{0.3cm}
    Ignacio Sanchez-Burgos$^{1}$, Pablo Montero de Hijes$^{2,3}$, Eduardo Sanz$^{4}$, Carlos Vega$^{4}$ and Jorge R. Espinosa$^{1,4*}$ \\ \vspace{0.15cm}
    \emph{$[1]$ Maxwell Centre, Cavendish Laboratory, Department of Physics, University of Cambridge, J J Thomson Avenue, Cambridge CB3 0HE, United Kingdom. \\
$[2]$ Faculty of Physics, University of Vienna, 1090 Vienna, Austria \\
$[3]$ Faculty of Earth Sciences, Geography and Astronomy, University of Vienna, 1090 Vienna, Austria \\ 
$[4]$ Department of Physical-Chemistry, Universidad Complutense de Madrid, Av. Complutense s/n, 28040, Madrid, Spain. \\
* = To whom correspondence should be sent.
email: jorgerene@ucm.es} \\

\end{center}
\thispagestyle{empty}

\section{Models details}
\subsection{Lennard-Jones potential}

We employ the Broughton and Gilmer Lennard-Jones potential ($u_{LJ-BG}$) \cite{broughton1986molecularAA}, which ensures that the potential and forces go to 0 at the cut-off distance:

\begin{equation}
u_{LJ-BG}=
\begin{cases} 4\epsilon\left[\left( \frac{\sigma}{r} \right)^{12}- \left( \frac{\sigma}{r}\right)^6 \right]+C_1 & \mbox{if }  r \leq 2.3\sigma \\
C_2\left(\frac{\sigma}{r}\right)^{12}+C_3\left(\frac{\sigma}{r}\right)^6+C_4\left(\frac{\sigma}{r}\right)^2+C_5 & \mbox{if } 2.3\sigma <r<2.5\sigma \\
0 & \mbox{if } 2.5\sigma\leq r \end{cases}
\label{PHS}
\end{equation}

where $C_1$=0.016132$\epsilon$, $C_2$=
3136.6$\epsilon$, $C_3$=
-68.069$\epsilon$, $C_4$=
-0.083312$\epsilon$,
and $C_5$=0.74689$\epsilon$. 

\subsection{mW potential}

The coarse-grained monoatomic water (mW) potential \cite{molinero2009waterAA} is based on the Stillinger-Weber (SW) silicon potential \cite{stillinger1985computerAA}, where tetrahedral coordination of the atoms is favored by adding  a three-body term that penalizes non-tetrahedral coordinations. This model is described by a potential ($u_{mW}$):

\begin{equation}
    u_{mW}=\phi_1(r_{ij})+\phi_2(r_{ij},r_{ik},\phi_{ijk})
\end{equation}

where $\phi_1$ is a standard pairwise potential and $\phi_2$ is the three-body term (dependent on the distance between atoms $i-j$, $i-k$ and the angle formed by the three of them. $\phi_1$ and $\phi_2$ are defined as:

\begin{equation} \black
    \phi_1(r_{ij})=A\epsilon_{mW}\left[B\left(\frac{\sigma}{r_{ij}}\right)^p-\left(\frac{\sigma}{r_{ij}}\right)^q\right]\mathrm{exp}\left(\frac{\sigma}{r_{ij}-a\sigma}\right)
\end{equation}

\begin{equation} \black
    \phi_2(r_{ij},r_{ik},\theta_{ijk})=\lambda\epsilon_{mW}[\mathrm{cos}\theta_{ijk} - \mathrm{cos}\theta_0]^2\mathrm{exp}\left(\frac{\gamma\sigma}{r_{ij}-a\sigma}\right)\mathrm{exp}\left(\frac{\gamma\sigma}{r_{ik}-a\sigma}\right)
\end{equation}

where $A=7.049556277$, $B=0.6022245584$, $p=4$, $q=0$, $\gamma=1.2$, $\theta_0=109.47$º, $\lambda=23.15$, $\epsilon_{mW}=6.189$ kcal/mol and $\sigma=2.3925$ \r{A}. The forces calculated using $u_{mW}$ vanish at a
cut-off distance of $r_c=a\sigma=1.8\ \sigma$.

\subsection{TIP4P/2005 potential}

The TIP4P/2005 water model \cite{abascal2005generalAA} is part of the TIP4P family of water models that extends the traditional three-point TIP3P model by adding an additional dummy atom (M, a massless site), where the charge associated with the oxygen atom is placed. In the TIP4P models there is one Lennard-Jones site for oxygen-oxygen interactions and three charged sites. The TIP4P/2005 model \cite{abascal2005generalAA} is one of the most successful models at describing the liquid–vapour coexistence properties, reproducing the experimental phase diagram \cite{vega2006vaporAA} and surface tension \cite{vega2007surfaceAA,mountain2009internallyAA,alejandre2010surfaceAA} reasonably well, as well as water transport properties \cite{abascal2005generalAA,vega2006vaporAA}.
\\

The interactions are described via a Lennard-Jones potential ($u_{LJ}$):

\begin{equation}
    u_{LJ}=4\epsilon_{LJ}\left[ \left( \frac{\sigma}{r}\right)^{12} - \left( \frac{\sigma}{r}\right)^6 \right]
    \label{ljequation}
\end{equation}

where $\epsilon_{LJ}$ accounts for the interaction strength; and a Coulombic interaction ($u_{Coulomb}$) for the charges:

\begin{equation}
    u_{Coulomb}=\frac{q_iq_j}{4\pi\epsilon_0r}
    \label{coulombeq}
\end{equation}

where $q_i$ and $q_j$ represent the site charge and $\epsilon_0$ the permittivity of vacuum. The parameters are summarized in Table \ref{parameters_tip4p}.
\\

\begin{table}[]
    \centering
    \begin{tabular}{c|c}
        Parameter & TIP4P/2005 \\ \hline
        $q_M/e$ & -1.1128 \\
        $q_H/e$ & 0.5564 \\
        $q_O/e$ & 0 \\
        $\epsilon_{lj}$ (O-O) / (kJ mol$^{-1}$) & 0.7749 \\
        $\sigma$ (O-O) / \r{A} & 3.1589 \\
        OH distance / \r{A} & 0.9572 \\
        $\theta$ of HOH angle & 104.52º \\
        OM distance / \r{A} & 0.1546 \vspace{0.3cm}
        
    \end{tabular}
    \caption{TIP4P/2005 model parameters.}
    \label{parameters_tip4p}
\end{table}

We simulate the TIP4P/2005 model with the GROMACS 4.6.7 MD package, and set the cut-off of both dispersion interactions and the real part of the electrostatic interactions at 12 Å. Long-range Coulombic interactions were treated with the Particle-Mesh Ewald (PME) solver \cite{darden1993particleAA,essmann1995smoothAA}. We kept the O–H bond length (0.9572 Å) and H–O–H angle (104.52$^o$) values constant with the LINCS algorithm \cite{hess1997lincsAA}.
\\

\section{Direct Coexistence simulations to determine the coexistence conditions}

We perform Direct Coexistence simulations with coexisting liquid and solid phases in order to determine the coexistence pressure at a given temperature \cite{Ladd1977Triple-pointSystemAA}. Once the simulation box is prepared with the two phases in contact, simulations at different pressures are performed. When the pressure is below the coexistence one, the fluid phase will grow at the expense of the solid phase, while the opposite behaviour occurs above the coexistence pressure. The conditions at which the simulations transition from crystallizing to melting is the coexistence pressure. This methodology can also be applied at constant pressure, and performing simulations at different temperatures. \textcolor{black}{With this method, one source of uncertainty comes from the difference in chemical potentials between the two phases outside the coexistence conditions. If they are too similar, multiple seeds can be necessary to narrow down the coexistence conditions accurately.}
\\

The barostat is applied only in the direction perpendicular to the interface. This requires ensuring that the simulation box can accommodate the solid phase at the exact conditions of the simulation i.e. the dimensions of the box in the non-elongated directions must yield a solid with a lattice with no stress \cite{Espinosa2013OnModelbAA}. \textcolor{black}{Therefore, for every pressure conditions evaluated, the dimensions of the simulation box and the atomic positions must be arranged accordingly. This and more potential complications on the determination of the coexistence conditions via direct coexistence simulations are discussed in Ref. \cite{frenkel2013simulationsAA}.}
\\

\textcolor{black}{For this study, the simulation boxes used were of the same size as those described in the main text to obtain $E$, $\sim$10000 particles. We previously determined for a different system that with this system size we were able to recapitulate the correct coexistence pressure \cite{sanchez2021fccAA}. Although the simulation length is only determined by the time it takes for the system to evolve into either the liquid or solid phase, this generally takes approximately t=1000 in reduced time units}.

\section{Mold Integration results}
\label{sec1}

As noted in the main text, we carry out Mold Integration calculations \cite{espinosa2014moldAA} in order to evaluate the crystal fluid interfacial free energy ($\gamma_{ls}$) for the (100) and (111) fcc Lennard-Jones planes. We do so by using the Mold LAMMPS package \cite{tejedor2024moldAA}, developed by us. The Mold Integration (MI) technique \cite{espinosa2014moldAA} is an accurate technique that allows the calculation of the interfacial free energy between a crystal and a fluid phase. The Gibbs free energy work, $\Delta G$, to reversibly form a crystal slab from the fluid can be directly related to the interfacial free energy as:

\begin{equation}
    \Delta G=2A\gamma_{ls}
    \label{gamma2}
\end{equation}

where 2A refers to the area of the two interfaces that the crystal slab exposes to the fluid. The formation of the crystal slab is induced by switching on an attractive interaction between a mold of potential energy wells and the fluid particles. The wells are placed at the equilibrium lattice positions of the crystal plane of interest at coexistence conditions. We model the particle-well interaction using a continuous square well (CSW) potential \cite{espinosa2014moldAA} ($u_{CSW}$) of the form:

\begin{equation} \black
u_{CSW}=-  \frac{1}{2} \epsilon_{CSW} \left[1-\mathrm{tanh}\left(\frac{r-r_w}{\alpha}\right) \right]
\label{CSW}
\end{equation}

where $\epsilon_{CSW}$ is the depth of the potential energy well, $r_w$ the radius of the attractive well, and $\alpha$ controls the steepness of the well. We choose $\alpha=0.005\sigma$.
\\

By gradually switching on the interaction between the potential wells and the fluid particles, we directly compute the required free energy work $\Delta G$ to induce the slab. Within this method, $\gamma$ is computed as follows: 

\begin{equation} \black
\gamma_{ls}(r_w)= \frac{1}{2A} (\epsilon_m N_w - \int_0^{\epsilon_m} d \epsilon_{CSW} ( \langle N_{fw}(\epsilon_{SCW})\rangle _{Np_xT})   \label{gamma2} 
\end{equation}

where $N_w$ is the total number of wells in the mold, and $<N_{fw}(\epsilon)>$ is the average number of filled wells in an $Np_xT$ simulation for potential wells of $\epsilon_{CSW}$ depth (the barostat in the simulation is only applied in the perpendicular direction to the crystal-fluid interface). The method consists in performing a thermodynamic integration along the path in which the depth of the potential mold wells is gradually increased to a maximum value of $\epsilon_{m}$. The integral of Eq. \ref{gamma2} must be reversible, and to ensure its reversibility, the crystal structure induced by the mold must quickly melt when the interaction between the potential wells and the fluid particles is switched off. To that end, the thermodynamic integration has to be performed at well radii $r_w$ that are wider than the optimal one, $r^o_w$. At $r_w=r_w^o$ the crystal slab is fully formed, and thus, its stability does not longer depend on the mold-fluid interactions. In practice, $\gamma_{ls}(r_w)$ can be estimated for several values of $r_w > r^o_w$, and then, be extrapolated to $r^o_w$, which is the well radius that provides the correct value of $\gamma_{ls}$. This technique has accurately obtained values of $\gamma_{ls}$ for ice---water interfaces of different models \cite{espinosa2016iceAA}, as well as fcc-liquid interfaces in Lennard-Jones and HS simulations \cite{espinosa2014moldAA}.
\\

\begin{figure}
    \centering
    \includegraphics[width=0.65\linewidth]{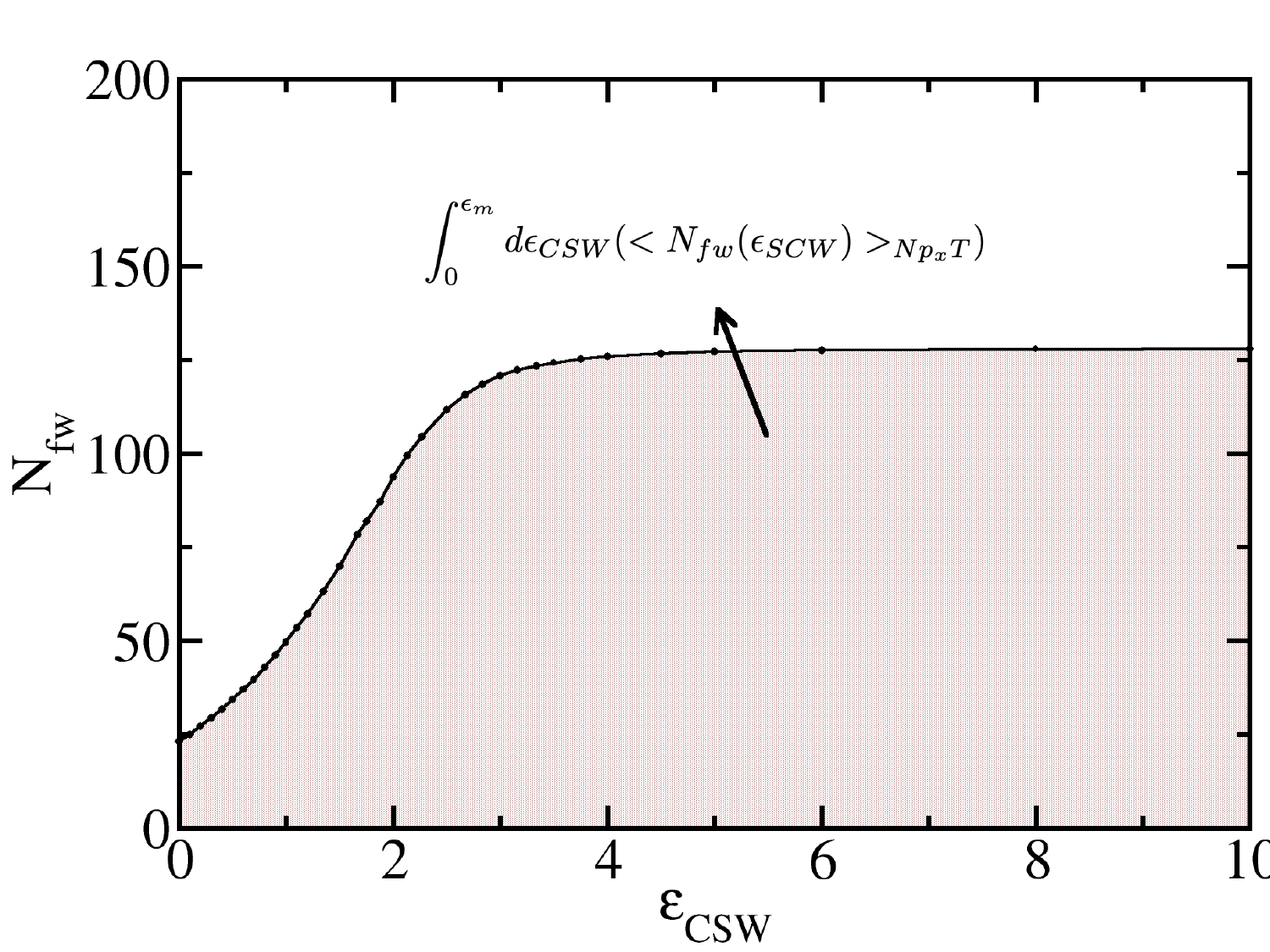}
    \caption{Number of filled wells (N$_{fw}$) as a function of the potential well depth of mold ($\epsilon_{CSW}$) for the Lennard-Jones system in the (100) crystal orientation at T$^*=$1 and $r_w=$0.36$\sigma$.}
    \label{fig_integral}
\end{figure}

\begin{figure}
    \centering
    \begin{tabular}{ccc}
        (a) $T^*=0.617$ & (b) $T^*=1$ & (c) $T^*=1.5$ \\
        \includegraphics[width=0.332\linewidth]{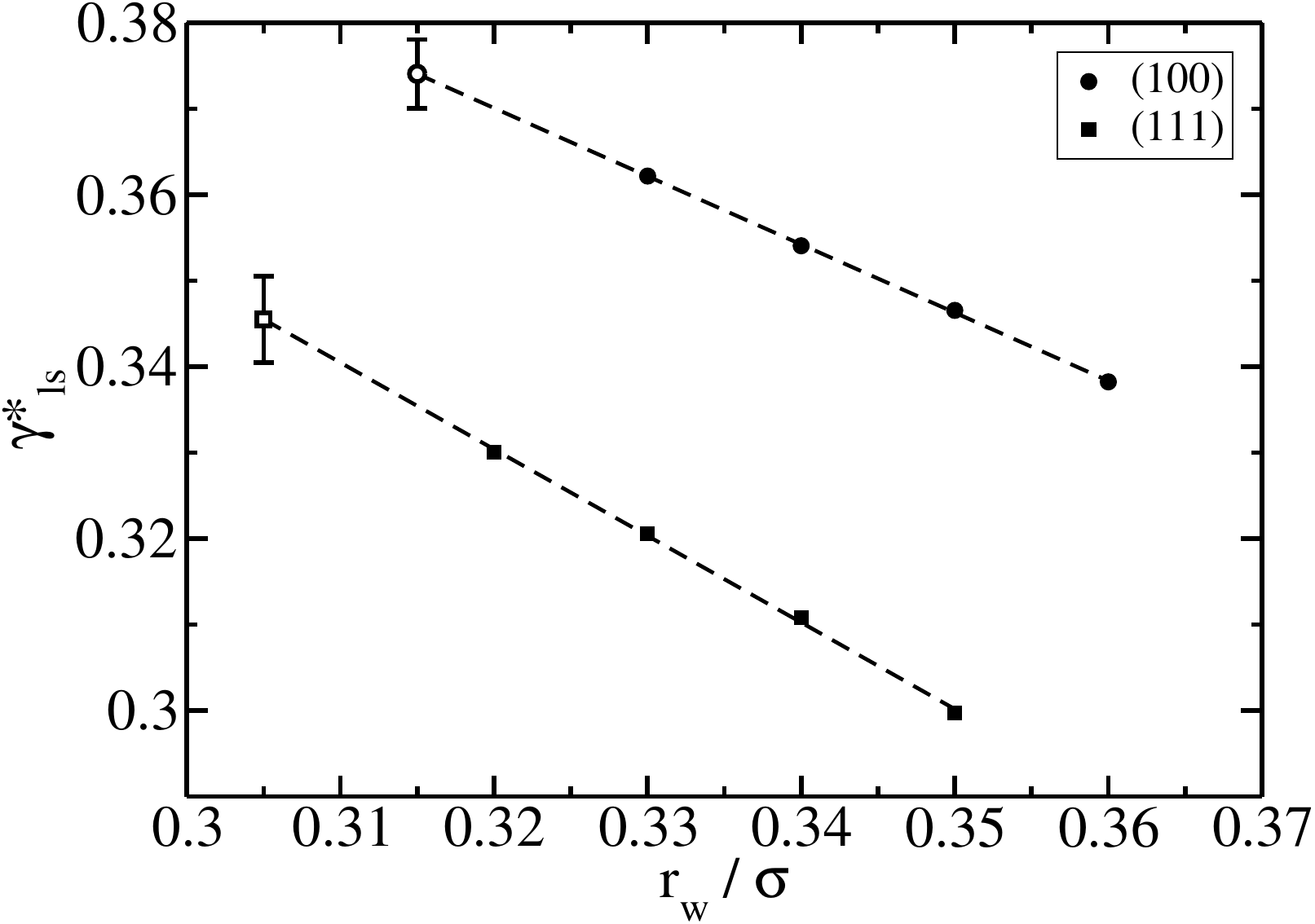} & \includegraphics[width=0.32\linewidth]{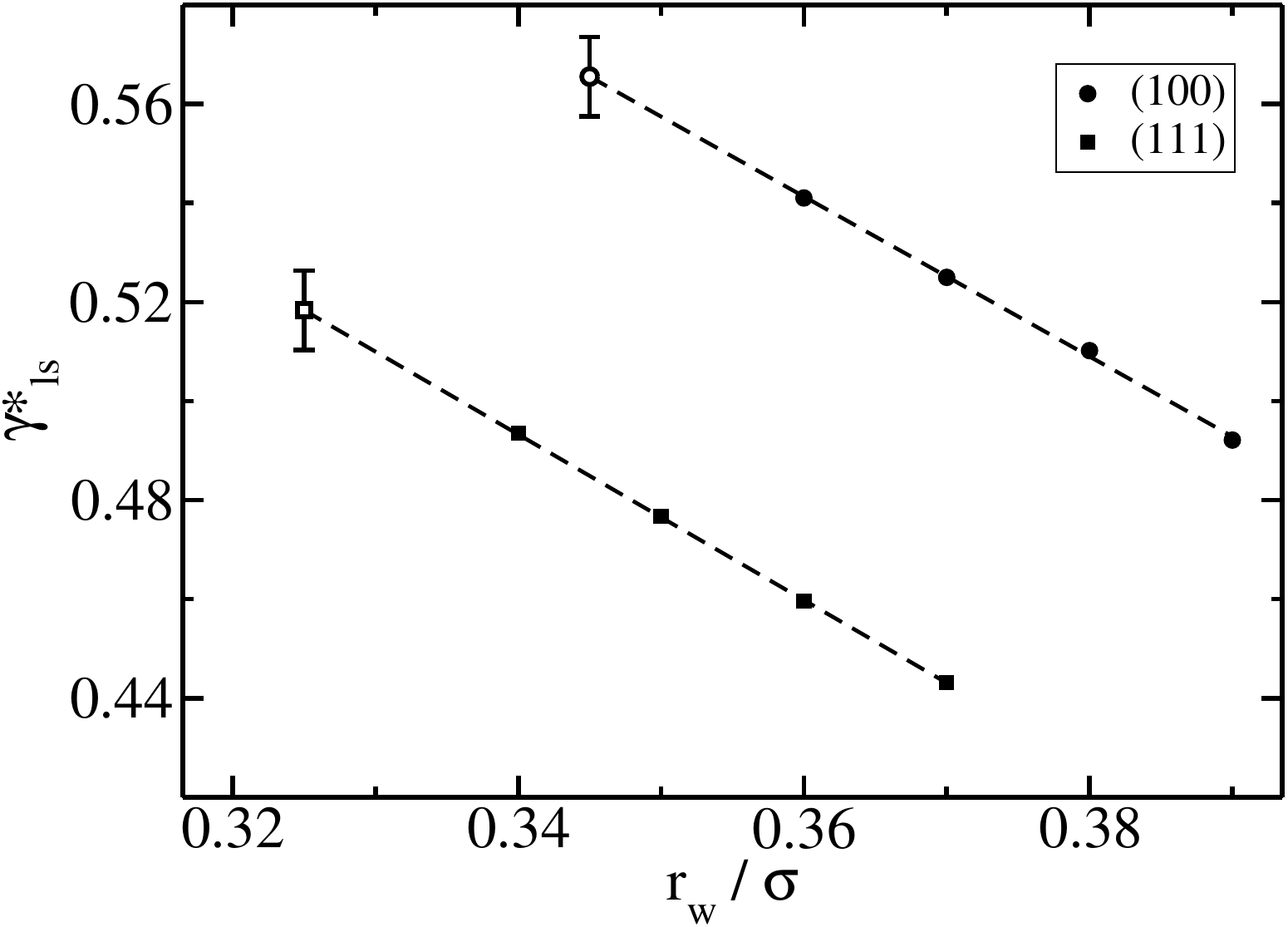} & \includegraphics[width=0.32\linewidth]{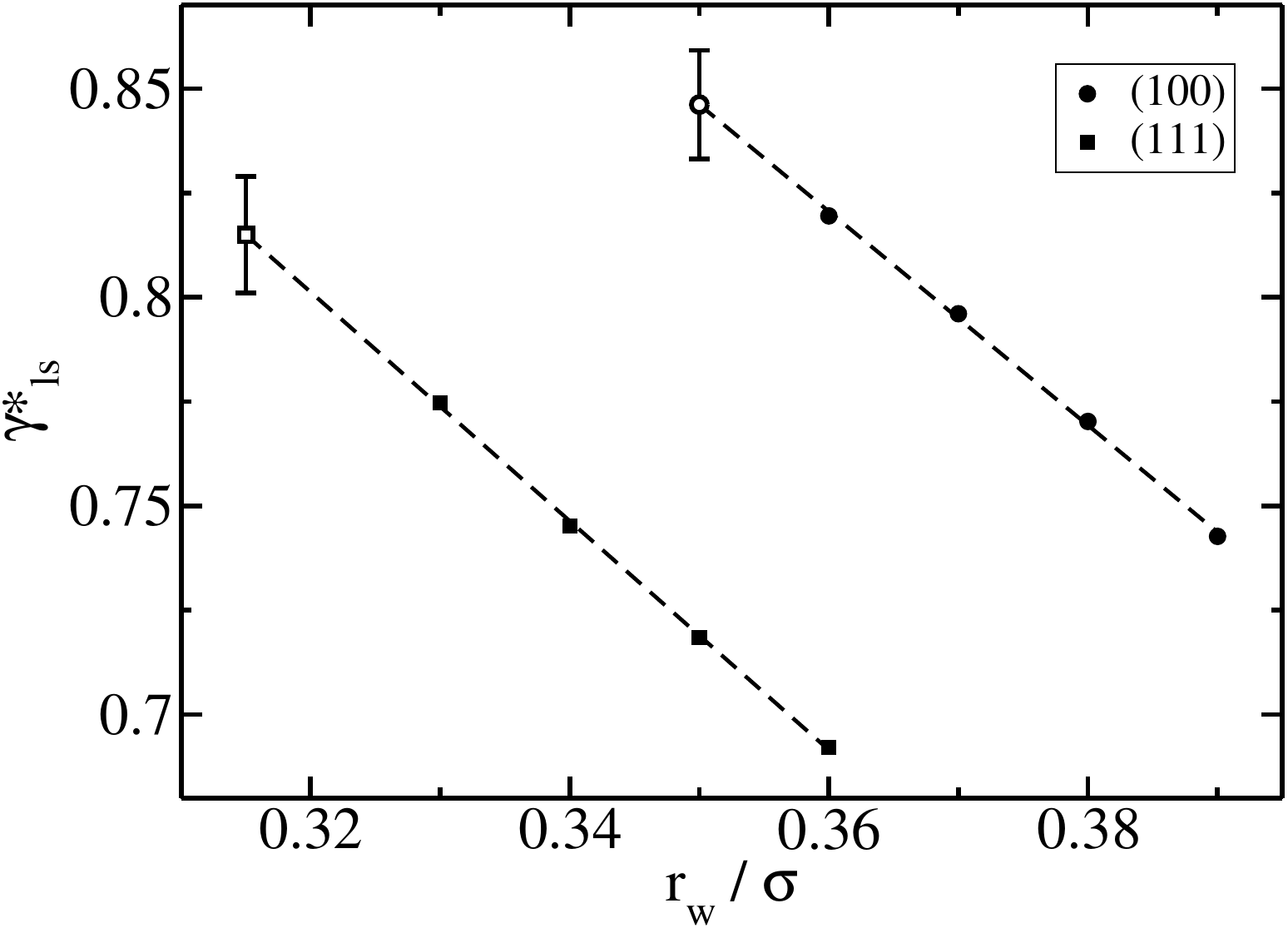}
    \end{tabular}
    \caption{\black (a)-(c) $\gamma_{ls}^*$ (where $\gamma_{ls}^*=\gamma_{ls}·(\epsilon\sigma^{-2})$) as a function of the potential well radius ($r_w$) for the different crystal orientations indicated in the figure legend. Filled symbols depict $\gamma_{ls}^*$ calculations at well widths wider than the optimal width. The extrapolation to the optimal well radius is represented with dashed lines, while the empty symbols depict the interfacial free energy at the optimal radius.}
    \label{fig_extrapolacion}
\end{figure}

In Figure \ref{fig_integral} we show the number of filled potential wells as a function of the potential depth ($\epsilon_{CSW}$) for the case of the (100) orientation at $T^*=1$ and $r_w=0.36\sigma$. The integral is obtained as the area below the curve, indicated as the shaded region in Figure \ref{fig_integral}. Then, $\gamma_{ls}$ is obtained through Eq. \ref{gamma2}. This integration is performed at radii wider than the optimal one, which we determined from independent simulations. Such optimal radius is determined as the largest well width that induces full crystallization of the system with no induction time (see Ref. \cite{espinosa2014moldAA}). For each different crystal orientation and temperature studied, we evaluated the optimal radius ($r_w^o$), and performed thermodynamic integration at radii wider than $r_w^o$. In Figure \ref{fig_extrapolacion} we plot the $\gamma_{ls}$ values obtained for the different cases (represented with filled symbols), as well as the extrapolation (dashed lines) to the optimal well radius (depicted with empty symbols), which gives the correct $\gamma_{ls}$ value. Moreover, we summarise the results from Mold Integration calculations in Table \ref{tab_mi}.

\begin{table}[]
    \centering
    \begin{tabular}{c|c|c|c}
        Plane & $T^*=0.617$ & $T^*=1$ & $T^*=1.5$ \\ \hline
        (100) & 0.374(4) & 0.566(8) & 0.846(13) \\ \hline
        (111) & 0.346(5) & 0.518(8) & 0.815(14)
    \end{tabular}
    \caption{Liquid-solid interfacial free energies ($\gamma_{ls}$) obtained for the Lennard Jones model from Mold Integration calculations.}
    \label{tab_mi}
\end{table}

\section{Thermodynamic integration results}

\subsection{Lennard-Jones model, liquid-solid interface equation of state}

\begin{figure}[h]
    \centering
    \includegraphics[width=0.55\linewidth]{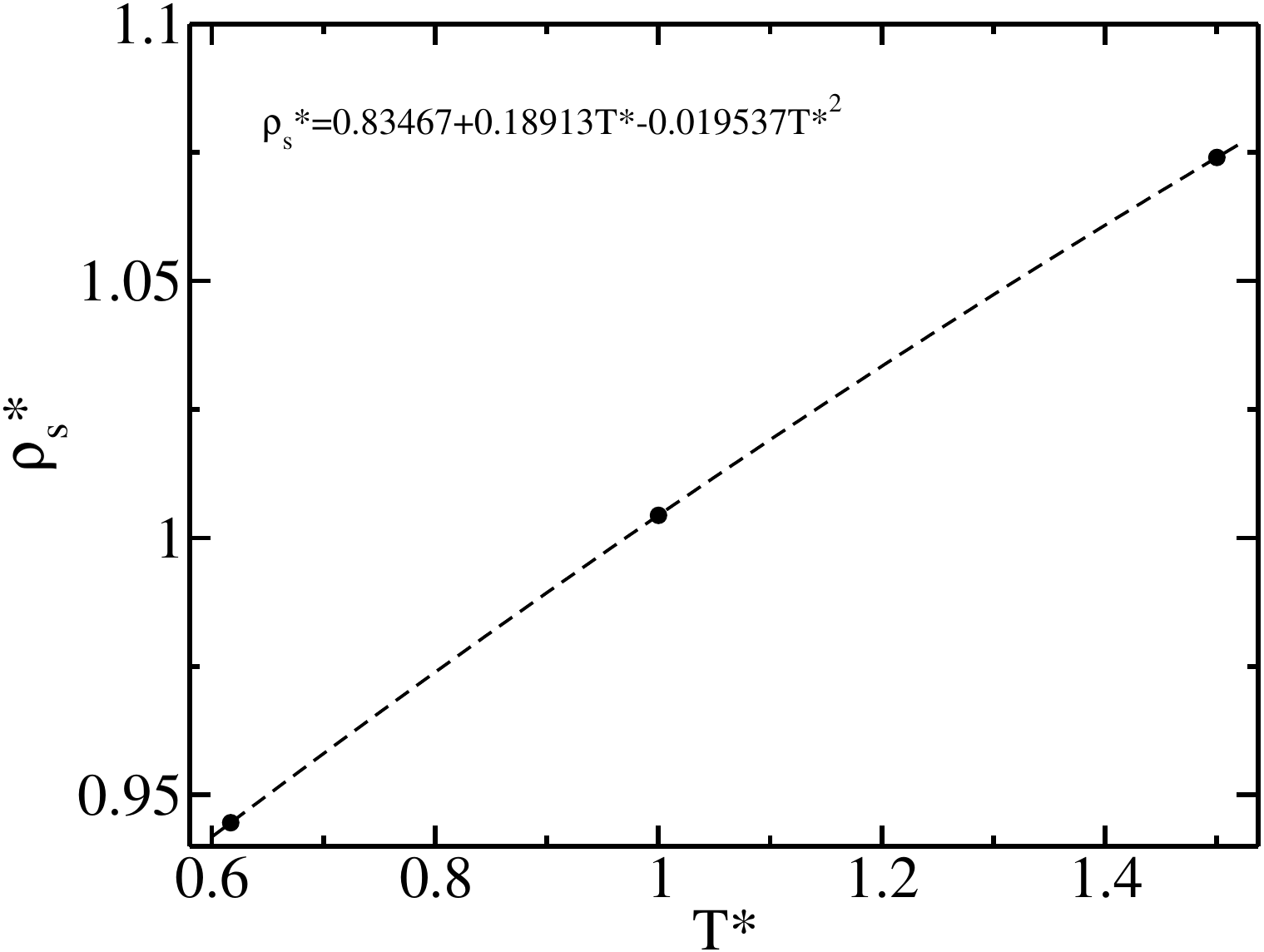}
    \caption{Equation of state ($\rho_s$ against $T^*$) of the fcc solid phase for the Lennard-Jones system along the coexistence line.}
    \label{fig_eos_lj}
\end{figure}

In Figure \ref{fig_eos_lj} we plot the Lennard-Jones fcc solid phase equation of state ($\rho_s$ against temperature), with the corresponding analytical expression fitted to the data.

\subsection{mW model equation of state}

\begin{figure}[h!]
    \centering
    \includegraphics[width=0.55\linewidth]{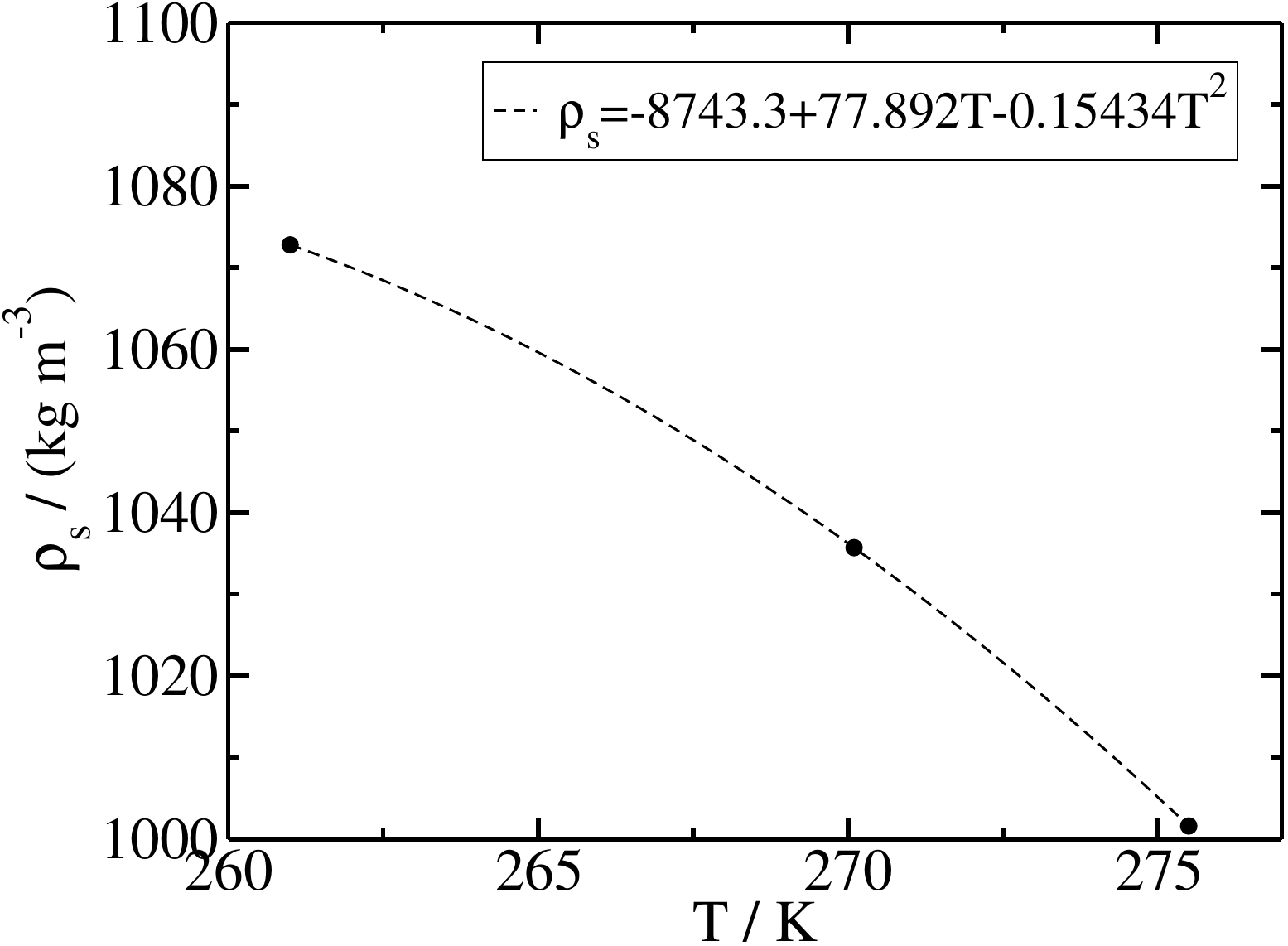}
    \caption{Equation of state ($\rho_s$ against $T$) of the Ice Ih solid phase for the mW water model along the coexistence line.}
    \label{fig_eos_mw}
\end{figure}

In Figure \ref{fig_eos_mw} we plot the mW ice Ih crystal phase equation of state ($\rho_s$ against temperature), with the corresponding analytical expression fitted to the data. \textcolor{black}{Please note that the solid density must be converted into number density in order to be used for calculations.}
\\

\newpage
\section{Thermodynamic integration equations}

     Laird et al. \cite{laird2009determinationAA} derived the main equation that we have applied in this work, being 

     \begin{equation}
    \left[\frac{d(\gamma_g/T)}{dT} \right]_{coex}=-\rho_s^{-2/3}\left[\frac{e}{T^2}+\frac{2\tau}{3\rho_sT}\left(\frac{d\rho_s}{dT}\right)_{coex}\right].
    \label{SMintegration}
\end{equation}

They find Eq. \ref{SMintegration} from this relation 

 \begin{equation}
     \frac{1}{A} d(\gamma A / T) = - \frac{e}{T^2}dT + \frac{\tau}{T}\frac{dA}{A},
     \label{eq:2}
 \end{equation}

by assuming cubic geometry 

\begin{equation}
    \rho_{s}^{2/3} = \frac{N_s^{2/3}}{A},
\end{equation}

and constant number of atoms in the crystal $N_s$. That is  

\begin{equation}
    \frac{2}{3}\rho_s^{-1/3} d\rho_s = - \frac{N_s^{2/3}}{A^2}dA = -\frac{\rho_s^{2/3}}{A}dA, 
\end{equation}

or   

\begin{equation}
    dA = - \frac{2A}{3\rho_s} d\rho_s.
 \end{equation}

 Therefore, we have,

 \begin{equation}
     \frac{1}{A} d(\gamma A / T) = - \left[ \frac{e}{T^2}dT + \frac{2\tau}{3\rho_s T} d\rho_s \right].
 \end{equation}

 Then, the Turnbull interfacial free energy per surface atom is used 
 
 \begin{equation}
     \gamma = \gamma_g \rho^{2/3},
 \end{equation}

so that 
\begin{equation}
    \frac{1}{A} d(\gamma A / T) = \frac{1}{A} d(\gamma_g \rho^{2/3} A / T)
\end{equation}

but note that $\rho^{2/3} A = N_s^{2/3}$ and

\begin{equation}
    \frac{1}{A} d(\gamma A / T) = \frac{1}{A} d(\gamma_g N_s^{2/3}  / T).
\end{equation}

Again, because $N_s$ does not change during this variation, then,

\begin{equation}
    \frac{1}{A} d(\gamma A / T) = \frac{N_s^{2/3} }{A} d(\gamma_g   / T) = \rho^{2/3}  d(\gamma_g   / T)
\end{equation}

and we find Eq. \ref{SMintegration}. Now, if we want
 to see what happens for fluid-fluid interfaces, we 
 can undo the change of variables and go back to the
  standard interfacial free energy, instead of using $\gamma_g$. This is,

 \begin{equation}
 \rho^{2/3} d(\gamma_g   / T)= \rho^{2/3} d(\gamma   / \rho_s^{2/3} / T) = d(\gamma/T) - \frac{2\gamma}{3T\rho_s}d \rho_s,
 \end{equation}
 which leads to 
   \begin{equation}
  d(\gamma  / T)= -  \frac{e}{T^2}dT -  (\tau - \gamma)\frac{2}{3\rho_s T} d\rho_s  
 \end{equation}

  and then the second term on the right-hand side is zero because for a fluid-fluid interface $\tau=\gamma$, i.e.

     \begin{equation}
  d(\gamma  / T)= -  \frac{e}{T^2}dT  
 \end{equation}

 \section{Results comparison}

\begin{figure}
    \centering
    \includegraphics[width=0.45\linewidth]{100_fits.pdf} \includegraphics[width=0.45\linewidth]{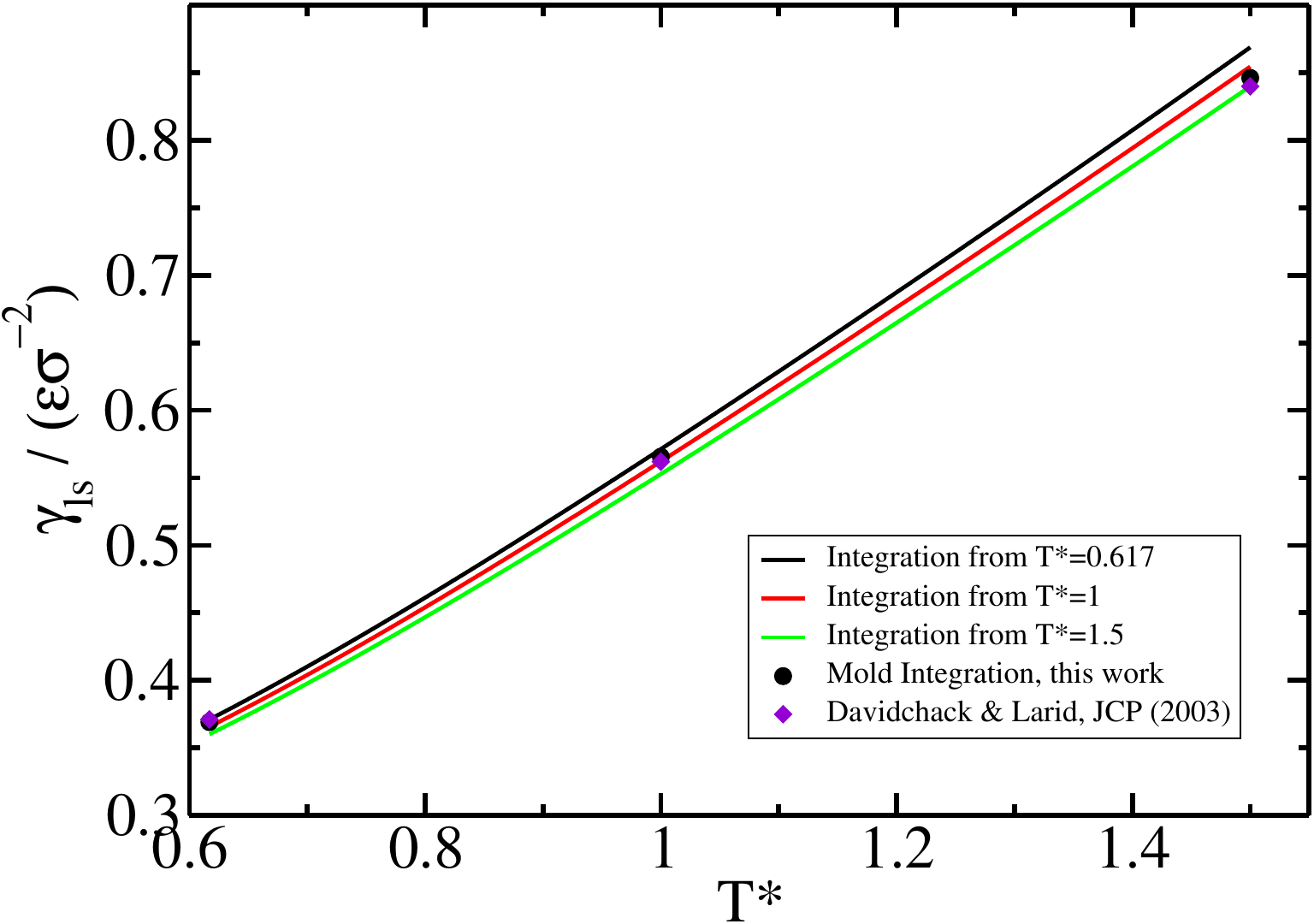}
    \caption{\black Interfacial free energy as predicted from Gibbs-Cahn integration for the LJ system and (100) crystal orientation, performed along the coexistence line. The $e$ and $\tau$ data used for this prediction was obtained: in this work and fitted the data with linear regression (left); and fitted to second order fits from the data by Laird \textit{et al.} \cite{laird2009determinationAA}.}
    \label{figcomp100}
\end{figure}

\begin{figure}
    \centering
    \includegraphics[width=0.45\linewidth]{110_fits.pdf} \includegraphics[width=0.45\linewidth]{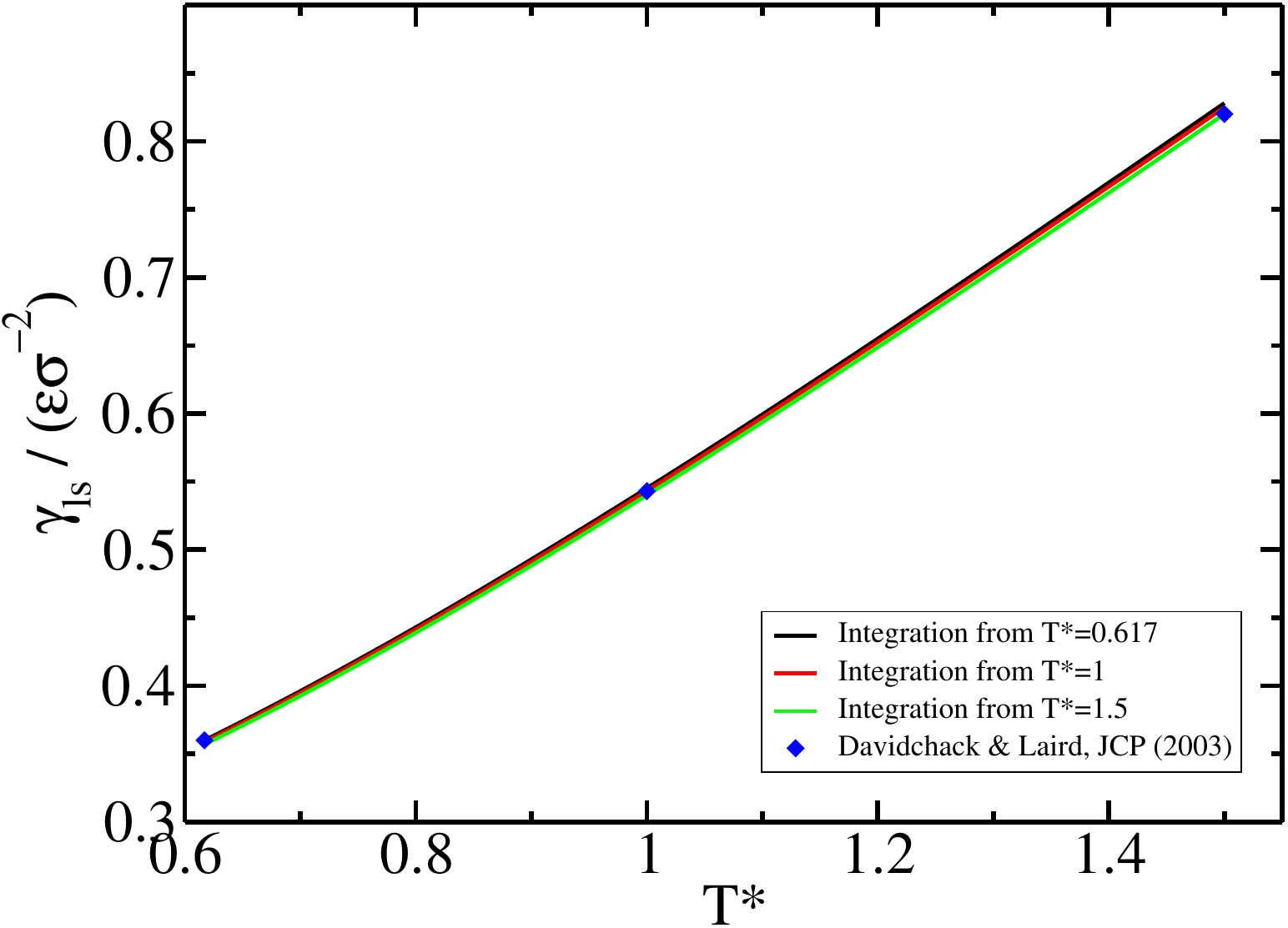}
    \caption{\black Interfacial free energy as predicted from Gibbs-Cahn integration for the LJ system and (110) crystal orientation, performed along the coexistence line. The $e$ and $\tau$ data used for this prediction was obtained: in this work and fitted the data with linear regression (left); and fitted to second order fits from the data by Laird \textit{et al.} \cite{laird2009determinationAA}.}
    \label{figcomp110}
\end{figure}

\begin{figure}
    \centering
    \includegraphics[width=0.45\linewidth]{111_fits.pdf} \includegraphics[width=0.45\linewidth]{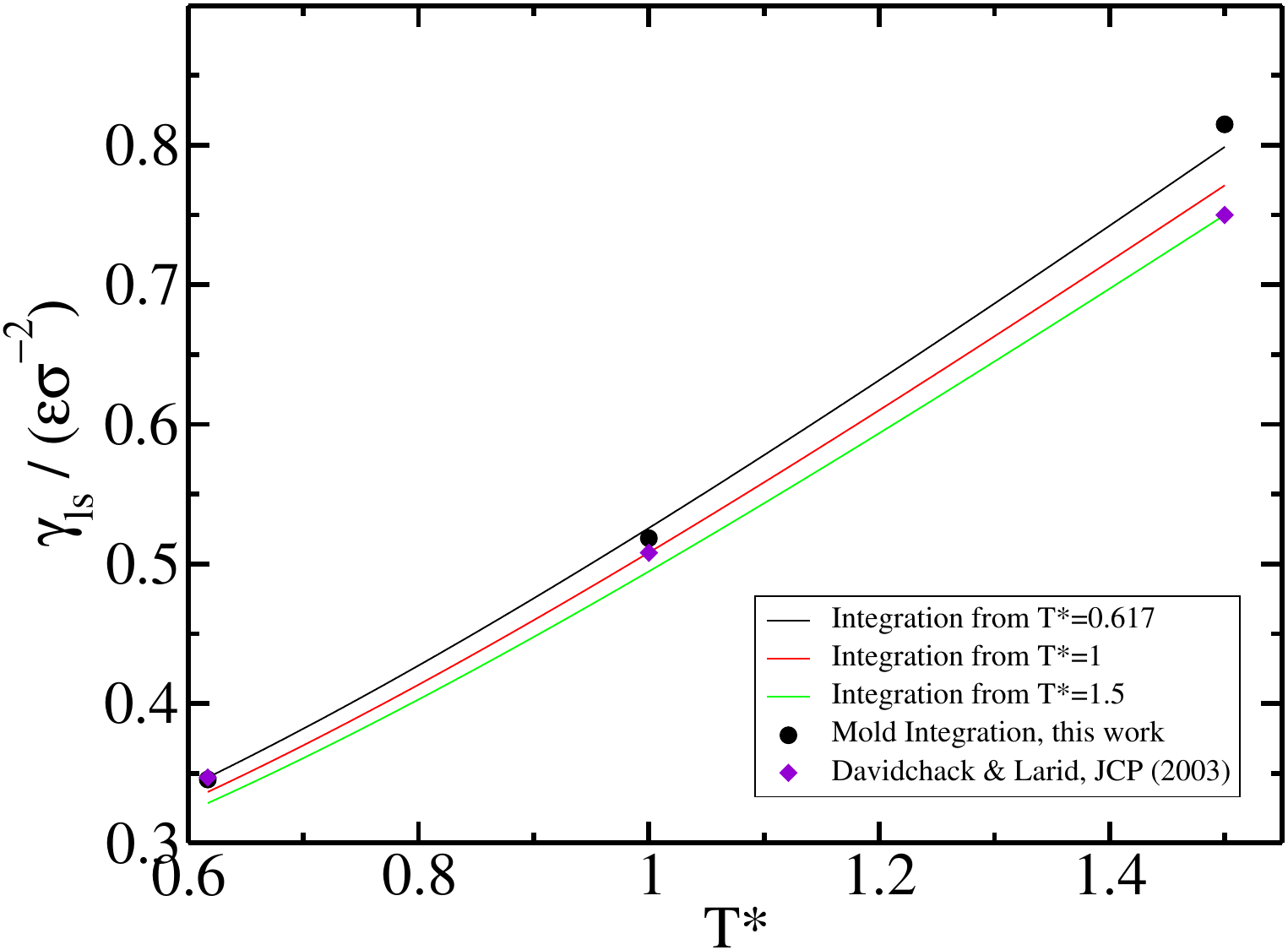}
    \caption{\black Interfacial free energy as predicted from Gibbs-Cahn integration for the LJ system and (111) crystal orientation, performed along the coexistence line. The $e$ and $\tau$ data used for this prediction was obtained: in this work and fitted the data with linear regression (left); and fitted to second order fits from the data by Laird \textit{et al.} \cite{laird2009determinationAA}.}
    \label{figcomp111}
\end{figure}

 \textcolor{black}{
 Here we show the predictions obtained from Gibbs-Cahn thermodynamic integration for the LJ system using either our data for $e$ and $\tau$ as a function of $T^*$ (as shown in the main text), and the results from Laird \textit{et al.} \cite{laird2009determinationAA}, using 2nd order fits to their data. In Figures \ref{figcomp100}, \ref{figcomp110} and \ref{figcomp111} we compare these two results for the (100), (110) and (111) orientations respectively.}

\bibliographystyle{ieeetr}

\end{document}